\documentclass[twoside]{article}

\usepackage[top=2cm,bottom=3cm, inner=3cm,outer=5.5cm]{geometry} 
\usepackage{nameref}
\usepackage[comma]{natbib}
\usepackage{url}
\usepackage{amssymb,amsmath}
\usepackage{bbm} 
\usepackage{tcolorbox} 
\usepackage{hyperref}

\newtcolorbox{textbox}[2][]{colback=blue!5!white, colframe=blue!75!black, fonttitle=\bfseries, title=#2, #1}

\usepackage{marginnote}
\usepackage[strict]{changepage} 
\newcommand{\oddentry}[2]{\marginpar{\setlength{\leftskip}{0.5cm}\small\textbf{#1:} #2}}
\newcommand{\evenentry}[2]{\marginpar{\setlength{\rightskip}{0.5cm}\small\textbf{#1:} #2}}
\newcommand{\entry}[2]{\checkoddpage\ifoddpage \oddentry{#1}{#2}\else \evenentry{#1}{#2}\fi}

\def\mbf#1{{
\mathchoice
{\hbox{\boldmath$\displaystyle{#1}$}}
{\hbox{\boldmath$\textstyle{#1}$}}
{\hbox{\boldmath$\scriptstyle{#1}$}}
{\hbox{\boldmath$\scriptscriptstyle{#1}$}}
}}
\def\vec{\mbf}

\newcommand{\E}{\mathbb{E}}
\newcommand{\cov}{\mathrm{cov}}
\newcommand{\var}{\mathrm{var}}
\newcommand{\Gau}{\mathrm{Gau}}
\newcommand{\tp}{{\!\scriptscriptstyle \top}} 
\newcommand{\norm}[1]{\left\lVert#1\right\rVert}

\DeclareMathOperator*{\argsup}{arg\,sup}

\def\d{\textrm{d}} 

\begin{document}

\sloppy

\title{Basis-Function Models in Spatial Statistics}
\author{Noel Cressie, Matthew Sainsbury-Dale, and Andrew Zammit-Mangion}
\date{\small School of Mathematics and Applied Statistics, University of Wollongong, Australia, NSW 2522; email: ncressie@uow.edu.au}
\maketitle

\begin{abstract}
Spatial statistics is concerned with the analysis of data that have spatial 
 locations associated with them, and those locations are used to model statistical dependence between the data. 
 The spatial data are treated as a single realisation from a probability model that encodes the dependence through both fixed effects and random effects, where randomness is manifest in the underlying spatial process and in the noisy, incomplete, measurement process. 
 The focus of this review article is on the use of basis functions to provide an extremely flexible and computationally efficient way to model spatial processes that are possibly highly non-stationary.
 Several examples of basis-function models are provided to illustrate how they are used in Gaussian, non-Gaussian, multivariate, and spatio-temporal settings, with applications in geophysics. Our aim is to emphasise the versatility of these spatial statistical models and to demonstrate that they are now centre-stage in a number of application domains. 
 The review concludes with a discussion and illustration of software currently available to fit spatial-basis-function models and implement spatial-statistical prediction.
 \end{abstract}


\section{Preface}\label{sec:preface}
 
 This article divides naturally into four principal sections.  
 Section \ref{part1} provides a succinct review of spatial statistics, its importance in scientific modelling and inference, particularly in the environmental sciences, and the role of basis-function models when the spatial data are `big'. 
 Section \ref{part2} presents Gaussian and non-Gaussian spatial-basis-function models, with particular emphasis on their role in hierarchical statistical models and spatial prediction. 
 Section \ref{part3} considers multivariate, warped, and spatio-temporal basis-function models. 
 The review concludes with Section \ref{part4}, which discusses software available, as of 2021, for spatial-basis-function modelling, and it uses a spatial data set of sea-surface temperatures for illustration.

\section{Introduction to spatial statistics and basis-function models}\label{part1}

 There are two main subsections in Section \ref{part1}, the first reviews spatial statistics and the second introduces basis-function models.
 
\subsection{Spatial statistics, grand challenges, and spatial stochastic processes}

In what follows in this subsection, we give a brief review of spatial statistics and its importance in scientific studies.

\subsubsection{Spatial statistics}\label{sec:SpatStatistics}

 Spatial statistics is concerned with the statistical analysis (both exploratory and confirmatory) of data indexed by spatial locations or regions in a spatial domain of interest. 
 A probabilistic framework is invoked that captures Tobler's famous `first law of geography' \citep{Tobler_1970_Toblers_Law}: ``Everything is related to everything else, but near things are more related than distant things.'' 
Statistical independence within a spatial data set is almost never assumed, although one should not rule out an eventual statistical model where spatial data are independent. 
 
 In spatial statistics, models have traditionally been proposed directly on the spatial data, but this modelling approach does not recognise a latent scientific process behind the noisy and incomplete (over the spatial domain) data. 
We shall see in Section \ref{sec:SpatStochasticProcceses} that ``what you see (data) is not what you want to get (process)'' \citep[][p.~xvi]{Cressie_Wikle_2011_stats_for_ST_data}.

\begin{textbox}{Grand challenges that need spatial statistics}
 Perhaps the greatest existential threat for humans is climate change; the planet is warming as its Homo sapiens are putting more and more greenhouse gases (in particular, carbon-based gases) into the atmosphere. 
The COP21 agreement signed in Paris in 2015 tasked each country to reduce their carbon emissions by an agreed-upon percentage by 2030, with the goal of capping global temperature increase at 2$^{\circ}$C above the pre-industrial level.  For a country to keep to its COP21 commitment, its government needs to know where the carbon sources are that can be decreased, and where the carbon sinks are that can be enhanced. These sources and sinks are more-or-less uncertain, and their estimation requires spatial-statistical methods \citep[e.g.,][]{Michalak_2004, Zammit_2021c}.

 Planet Earth also faces a resources `grand challenge' of producing enough food, water, energy, and shelter for its inhabitants. Equally important for Earth's future, we Homo sapiens must collectively recognise that the health of its diverse species is critical to our own health. 
 For example, not only does logging of the Amazon region move carbon from the biosphere into the atmosphere where it is dangerous, it destroys habitat for plants and animals, severely impacting an  ecosystem that is a source of food and medicine for our species. 
 Spatial data sets collected by ecologists are at the centre of monitoring and assessment of critical ecosystem health \citep[e.g.,][]{Hooten_2017_Animal_movement_telemetry_data}.
\end{textbox}
 
 \subsubsection{Spatial stochastic processes and their measurement}\label{sec:SpatStochasticProcceses}

 Models are a way to focus on the important aspects of a problem, and in this subsection we start with a univariate process of interest, defined on a domain $D$ that is a subset of $d$-dimensional Euclidean space.
 A spatial stochastic process on $D$ is written as \entry{Spatial stochastic process}{A countable or uncountable set of random variables indexed by locations in a spatial domain.}
 \begin{equation}\label{eqn:ProcessModel}
 Y(\cdot) \equiv \{Y(\vec{s}): \vec{s} \in D\},
 \end{equation}
where $D \subset \mathbb{R}^d$ and, at $\vec{s} \equiv (s_1, \dots, s_d)^\tp$, $Y(\vec{s})$ is a random quantity. On $D$, we define the Euclidean distance, $\norm{\vec{s} - \vec{u}} \equiv \{\sum_{i=1}^d(s_i - u_i)^2\}^{1/2}$, between spatial locations $\vec{s}$ and $\vec{u}$ in $D$.  
 The process $Y(\cdot)$ is well defined if all finite probability distributions of $(Y(\vec{s}_1), \dots, Y(\vec{s}_m))^\tp$ exist for all $\{\vec{s}_1, \dots, \vec{s}_m\} \subset D$ and all $m \geq 1$, and they satisfy Kolmogorov's consistency conditions \citep[e.g.,][Sec.~2.4]{Tao_2011_Measure_theory}. \citet[Ch.~1]{Cressie_1993_stats_for_spatial_data} and \citet[Ch.~1]{Banerjee_2004} give three main cases for $D$, resulting in geostatistical processes ($D$ fixed, uncountable with Lebesgue measure $|D| > 0$), lattice spatial processes ($D$ fixed, countable), and marked point processes ($D$ a countable subset of a fixed domain). 
 See also \citet[][Sec.~1.1]{Cressie_1993_stats_for_spatial_data} for an integrative spatial model covering all these cases, where \textit{both} $Y(\cdot)$ and $D$ in Equation \ref{eqn:ProcessModel} are assumed random. 

 In the rest of this review, we concentrate on basis-function models for the geostatistical case. 
 However, basis-function models certainly have a place in spatial statistics for lattice data \citep[e.g.,][]{Bradley_2016_Bayesian_spatial_COS_lattice_data, Bradley_2020_Bayesian_Hierarchical_Conjugate_Full_Conditional_Distributions_Natural_Exponential_Family} and point patterns 
 (e.g., \citeauthor{Cseke_2016} \citeyear{Cseke_2016}, \citeauthor{Simpson_2016} \citeyear{Simpson_2016}, \citeauthor{Hooten_2017_Animal_movement_telemetry_data} \citeyear{Hooten_2017_Animal_movement_telemetry_data}, Ch.~4).  
 
 To introduce the spatial data $\vec{Z} \equiv (Z(\vec{s}_1), \dots, Z(\vec{s}_n))^\tp$ into the probability model, we specify the \textit{data model}, $[\vec{Z} \mid Y(\cdot)]$, which is the probability distribution of $\vec{Z}$ given $Y(\cdot)$. 
 Then the joint distribution of the latent spatial process $Y(\cdot)$ and the spatial data $\vec{Z}$ is
 \begin{equation}\label{eqn:JointDistribution}
 [Y(\cdot), \vec{Z}] = [\vec{Z} \mid Y(\cdot)] [Y(\cdot)], 
 \end{equation}
 where henceforth the probability model $[Y(\cdot)]$ of the latent process $Y(\cdot)$ given by Equation \ref{eqn:ProcessModel}, is referred to as the \textit{process model}. 
 In Equation \ref{eqn:JointDistribution} and elsewhere, the bracket notation is used: For generic random quantities $A$ and $B$, $[A, B]$ denotes the joint probability distribution of $A$ and $B$; $[A \mid B]$ denotes the conditional probability distribution of $A$ given $B$; and $[A]$ denotes the marginal probability distribution of $A$.
 
 As explained in Section \ref{sec:SpatStatistics}, the data $\vec{Z}$ are incomplete (i.e., do not cover all of the spatial domain $D$) and noisy (i.e., contain measurement error), which is captured in the data model, $[\vec{Z} \mid Y(\cdot)]$. 
 A common assumption made in the data model is that of conditional independence, since usually the act of measurement made at one location is statistically independent of the act of measurement at another. 
 Henceforth in this review, the data model is assumed to exhibit this conditional independence, namely 
  \begin{equation}\label{eqn:DataModel}
 [\vec{Z} \mid Y(\cdot)] = \prod_{i=1}^n [Z(\vec{s}_i) \mid Y(\vec{s}_i)]. 
 \end{equation}
 
 The spatial dependence one expects from `Tobler's first law' is captured probabilistically in the process model $[Y(\cdot)]$. 
 Then the spatial-statistical dependence in the spatial data $\vec{Z}$ that Tobler articulated is inherited from $[Y(\cdot)]$ through the expression, 
 \begin{equation*}
 [\vec{Z}] = \int [\vec{Z} \mid Y(\cdot)] [Y(\cdot)] \d Y(\cdot), 
 \end{equation*}
where the integral is used formally to represent marginalisation of the joint probability measure of $Y(\cdot)$ and $\vec{Z}$. 
The combination of the probability model $[Y(\cdot)]$ and Equation \ref{eqn:DataModel} defines a \textit{hierarchical spatial-statistical model} \citep[e.g.,][]{Wikle_Berliner_2007_Bayesian_tutorial}. 

 In this review, our principal interest is in spatial prediction of $Y(\vec{s}_0)$ at a given spatial location $\vec{s}_0 \in D$. 
 To do this, we use the predictive distribution, $[Y(\vec{s}_0) \mid \vec{Z}]$, which is given by Bayes' Rule, 
  \entry{Bayes' Rule}{For generic random quantities $A$ and $B$, Bayes' Rule relies on the simple relationship, $[A \mid B]$ $=$ $[B \mid A][A]/ [B]$. 
 }
  \begin{equation}
 [Y(\vec{s}_0) \mid \vec{Z}] = [\vec{Z} \mid Y(\vec{s}_0)][Y(\vec{s}_0)]/[\vec{Z}].\label{eqn:predictiveDistribution}
  \end{equation}
 When both the data $\vec{Z}$ and the process $Y(\cdot)$ are Gaussian, this predictive distribution can be computed using standard Gaussian identities (see Section~\ref{sec:Basis-function representations in Gaussian spatial processes}). When the data and/or the process model are non-Gaussian, approximations or Monte Carlo sampling-based techniques are typically needed for inferences such as spatial prediction (see Section \ref{sec:Basis-function representations in non-Gaussian spatial processes}).

\subsection{Basis-function representations of spatial processes and covariance functions}\label{sec:Basis-function representations of spatial processes and covariance functions}

In the subsections that follow, we use basis functions to represent spatial processes and their corresponding spatial covariance functions.

\subsubsection{Basis-function representations of spatial processes}\label{sec:Basis-function representations of spatial processes}

 From a mathematical point of view, a basis of a function space is a collection of elements for which any function in the space can be represented as a linear combination of these basis elements. 
Relevant to this review, a countable basis $\{\phi_j (\cdot): j = 1, 2, \dots\}$ of the space of square-integrable functions, $\mathcal{F}$, has the property that any function $f(\cdot) \in \mathcal{F}$ can be represented as 
 \begin{equation*}
 f(\cdot) = \sum_{j \geq 1} a_j \phi_j (\cdot), 
 \end{equation*}
 in an $L_2$ sense. 
 \entry{Inner product}{The inner-product operation for two elements $f(\cdot)$ and $g(\cdot)$ of a function space whose functions are square integrable on the domain $D$, is $\int_D f(\vec{s}) g(\vec{s}) \d \vec{s}$.}
 \entry{Orthogonality}{$f(\cdot)$ and $g(\cdot)$ are \textit{orthogonal} if their inner product is zero.}
 \entry{Orthonormality}{$f(\cdot)$ and $g(\cdot)$ are \textit{orthonormal} if they are orthogonal and the inner products of $f(\cdot)$ and $g(\cdot)$ with themselves are both equal to 1.}
 Furthermore, if the elements of $\{\phi_j (\cdot): j = 1, 2, \dots\}$ are mutually orthonormal in terms of the inner product, then, straightforwardly, $a_j = \int_D f(\vec{s}) \phi_j(\vec{s}) \d \vec{s}$ for $j \geq 1$. Orthogonality of a basis is attractive but not necessary for a good representation of a function's behaviour.  For example, spline bases are not usually orthogonal and, while many wavelet bases are, there may be computational reasons for using non-orthogonal wavelets.

 Now we switch our point of view from mathematical to statistical: Consider a probability measure on the space of spatial functions on $D$ defined by a finite random linear combination of basis functions plus an independent error term \citep[e.g.,][]{Solo_2002_stochastic_heat_equation_EM_algorithm}. 
 The probability measure comes from the representation 
 \begin{equation}
 Y(\cdot) = \sum_{j = 1}^r \alpha_j \phi_j(\cdot) + \delta(\cdot),\label{eqn:SREmodel}
 \end{equation}
 where $\sum_{j = 1}^r \alpha_j \phi_j(\cdot)$ is the basis-function term consisting of a linear combination of the basis functions $\{\phi_j (\cdot): j = 1, \dots, r\}$; the coefficients $\{\alpha_j: j = 1, \dots, r\}$ are in general dependent random variables; and the error term, $\delta(\cdot) \equiv \{\delta(\vec{s}) : \vec{s} \in D\}$, is a stochastic process independent of the basis-function coefficients.  
 The inclusion of $\delta(\cdot)$ in Equation \ref{eqn:SREmodel} accounts statistically for the error incurred by using only a finite number of basis functions. As we discuss below, it preserves variability of $Y(\cdot)$ and, in that sense, Equation \ref{eqn:SREmodel} is a statistical model, not an approximation, of spatial variability.

 The basis functions in Equation \ref{eqn:SREmodel} may be defined by truncations of a countable basis, or they may simply be functions thought to be important for representing the spatial variability. 
 For example, basis functions can come in the form of splines \citep{Wahba_1990_spline_models_for_observational_data}, wavelets \citep{Vidakovic_1999_introduction_to_wavelets}, bisquare functions \citep{Cressie_Johannesson_2008_FRK}, Wendland functions \citep{Nychka_2015}, and finite elements \citep{Lindgren_2011}. The basis functions of \citet{Banerjee_2008} depend on covariance-function parameters (unlike the examples given above), which  results in slower computations when fitting and predicting \citep{Bradley_2016_comparison_of_spatial_predictors}. 
 The resulting process model $[Y(\cdot)]$, where $Y(\cdot)$ is given by Equation \ref{eqn:SREmodel}, has been called the spatial random effects (SRE) model \citep{Cressie_Johannesson_2008_FRK}, which is generalised below to a spatial mixed effects model and used extensively in spatial prediction from very large data sets \citep{Zammit_2021a}. 
 A review of such low-rank representations can be found in \cite{Wikle_2010_low_rank_representations_for_spatial_processes}.  
 
 In the hierarchical model (HM), $Y(\cdot)$ represents the scientific process, and hence $\{\var(Y(\vec{s})):\vec{s} \in D\}$ should be conserved, no matter what basis-function representation is chosen. 
   Equation \ref{eqn:SREmodel} conserves variability in that  $\var(Y(\vec{s})) = \var\big(\sum_{j = 1}^r \alpha_j \phi_j(\vec{s})\big) + \var(\delta(\vec{s}))$ \citep{Cressie_Johannesson_2008_FRK}. 
 That is, Equation \ref{eqn:SREmodel} is a \emph{model}, not an approximation for $Y(\cdot)$. 
 Clearly, $\delta(\cdot)$ is a critical component, since $\var(Y(\vec{s})) \geq \var(\sum_{j = 1}^r \alpha_j \phi_j(\vec{s}))$ for all $\vec{s} \in D$. 
 Of course, the model for $\delta(\cdot)$ changes if the basis-function component changes, since the variances and covariances of the total, $Y(\cdot)$, are fixed by the underlying science. 
 
 Throughout this article, the basis functions are known, and the randomness is in their coefficients $\{\alpha_j: j = 1, \dots, r\}$. 
 There is another type of representation we shall mention briefly, known as spatial factor analysis \citep{Christensen_Amemiya_2002_Latent_Variable_Analysis_of_Multivariate_Spatial_Data, Christensen_Amemiya_2003_multivariate_spatial_factor_analysis}. 
 The models are defined for multivariate spatial processes, where the factors are spatial functions that are random, and the coefficients are deterministic but have to satisfy critical identifiability conditions; further discussion is given in Sections \ref{sec:multivariate} and \ref{sec:ST_dynamic}.

\begin{figure}
  \includegraphics[width=\linewidth]{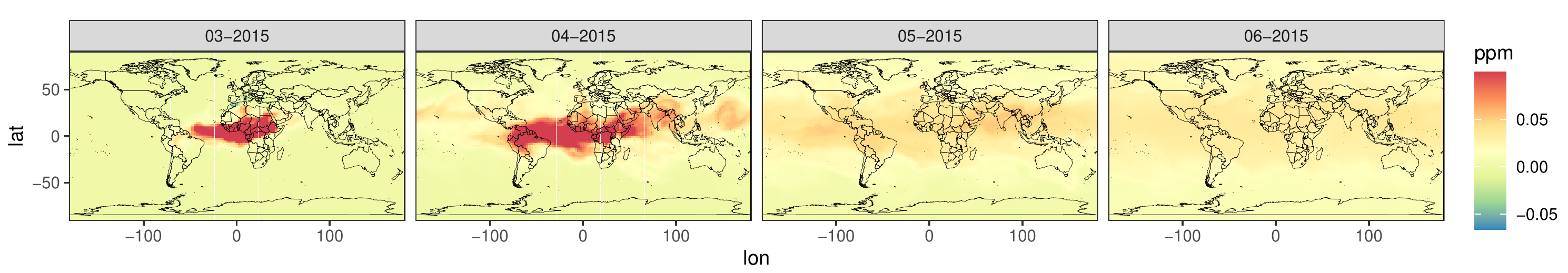}
  \caption{A spatio-temporal basis function, in units of parts per million (ppm) of $\mathrm{CO_2}$, generated using the transport model MOZART in response to a pulse of $\mathrm{CO_2}$ flux in Northern Africa throughout the month of March 2015. The plots show the $\mathrm{CO_2}$ concentration in ppm on March 15, April 15, May 15, and June 15, 2015, respectively.\label{fig:mfresponse}}
\end{figure}

 Now basis functions are often multi-resolutional (e.g., wavelets, bisquares) but some of them could be physical. 
 A physical basis function could be some readily available geophysical quantity, such as the elevation of the terrain as a function of $\vec{s}\in D$ when modelling (say) maximum temperature in a geographic region $D$. 
 In more complicated cases, a basis function could be the output of a numerical model, such as an atmospheric-transport model. For example, \textbf{Figure~\ref{fig:mfresponse}} summarises the three-dimensional spatio-temporal output of the Model for OZone And Related chemical Tracers (MOZART), a numerical atmospheric-transport model. The output is simulated CO$_2$ changes in the atmosphere in response to a pulse of emissions in central Africa throughout the month of March 2015. Several of these basis functions can be constructed by `driving' the atmospheric model with pulses at different times and in different regions, which can then be used to model a temporal sequence of spatial fields of $\mathrm{CO_2}$ (in parts per million).  This approach to basis-function modelling of CO$_2$ is ubiquitous in the atmospheric sciences \citep[e.g.,][]{Enting_2002}, and it was developed into a fully Bayesian statistical framework by \cite{Zammit_2021c}. 
 Another example of physically motivated basis functions was provided by  \cite{Wikle_2001_spatiotemporal_tropical_ocean_surface_winds}, who used the equatorial normal mode orthogonal basis functions to model tropical ocean surface winds.

 The general basis-function model with non-zero mean $\mu(\cdot)$ can be written as
 \begin{equation}
 Y(\cdot)
 = \mu(\cdot) + \sum_{j = 1}^r \alpha_j \phi_j(\cdot) + \delta(\cdot)
 = \vec{x}(\cdot)^\tp \vec{\beta} + \vec{\phi}(\cdot)^\tp \vec{\alpha} + \delta(\cdot),\label{eqn:general_basis-function_model}
  \end{equation}
 where the second equality is imposed when the mean is expressed as a linear regression on a $p$-dimensional vector of covariates $\vec{x}(\cdot)$; that is, $\mu(\cdot) = \vec{x}(\cdot)^\tp \vec{\beta}$.  
 Equation \ref{eqn:general_basis-function_model} is called a \textit{spatial mixed effects model}, since now the fixed effects $\vec{x}(\cdot)^\tp \vec{\beta}$ are added to the spatial random effects expressed in Equation \ref{eqn:SREmodel}.
 In Equation \ref{eqn:general_basis-function_model}, we have written the linear combination of basis functions as $\sum_{j = 1}^r \alpha_j \phi_j(\vec{\cdot}) = \vec{\phi}(\cdot)^\tp \vec{\alpha}$, where $\vec{\phi}(\cdot) \equiv (\phi_1(\cdot), \dots, \phi_r(\cdot))^\tp$ and $\vec{\alpha} \equiv (\alpha_1, \dots, \alpha_r)^\tp$. 
 Functions  such as a constant or trends in the components of $\vec{s}$, are clearly covariates, but physical functions such as elevation could be a covariate (fixed effect) or a basis function (random effect), depending on the application. 
 Basis functions whose coefficients are deterministic are part of the fixed effects and are considered as covariates; overfitting of the fixed effects can be handled by regularization, such as with the least absolute shrinkage and selection operator \citep[lasso;][]{Tibshirani_1996_lasso}, and this has a Bayesian interpretation.

 In what follows, Equation \ref{eqn:general_basis-function_model} will be the process model in the HM.  
 Data $\vec{Z} = (Z(\vec{s}_1), \dots, Z(\vec{s}_n))^\tp$ are an imperfect `look' at $Y(\cdot)$, and the principal inferential goal is to predict $Y(\cdot)$ over $D$ with known or model-derived (in practice, estimated) uncertainty. 
 
\subsubsection{Spatial covariance functions for basis-function models}\label{sec:covariance_functions}

\entry{Positive-semidefinite}{A covariance function $C(\cdot, \cdot)$ is positive-semidefinite if $\sum_{i = 1}^m \sum_{j = 1}^m$ $C(\vec{s}_{i}, \vec{s}_{j}) a_{i}\bar{a}_{j} \geq 0$ for any $\vec{s}_{i} \in \mathbb{R}^d$,  $i = 1, \dots, m$, any complex $\{a_{i}: i = 1, \dots, m\}$, and any integer $m > 0$. (Here, $\bar{a}_{j}$ is the complex conjugate of $a_j$.)}
 The covariance function of $Y(\cdot)$ is defined as
 \begin{equation}\label{eqn:cov_fun}
 C_Y(\vec{s}, \vec{u}) \equiv \cov(Y(\vec{s}), Y(\vec{u})),\quad \vec{s}, \vec{u} \in D.
 \end{equation}
 The covariance function is a key functional parameter in spatial statistics, as it serves as the primary mechanism for capturing the dependence one expects from `Tobler's first law'. 
  For a spatial-statistical model to be valid, the covariance function is required to be positive-semidefinite. 
  Recalling that the term $\delta(\cdot)$ is a stochastic process independent of the basis-function coefficients $\vec{\alpha}$, the covariance function implied by Equation \ref{eqn:general_basis-function_model} is
 \begin{align}
 C_Y(\vec{s}, \vec{u}) 
 &= \cov(\vec{x}(\vec{s})^\tp \vec{\beta} + \vec{\phi}(\vec{s})^\tp \vec{\alpha} + \delta(\vec{s}), \; \vec{x}(\vec{u})^\tp \vec{\beta} + \vec{\phi}(\vec{u})^\tp \vec{\alpha} + \delta(\vec{u})) \nonumber\\
 &= \vec{\phi}(\vec{s})^\tp\vec{K}\vec{\phi}(\vec{u}) +  
 C_\delta(\vec{s},\vec{u}),
 \label{eqn:C(s, u)}
 \end{align}
 where $\vec{K} \equiv \cov(\vec{\alpha}, \vec{\alpha})$, and $C_\delta(\vec{s},\vec{u})$ is (usually) a simple covariance function that describes fine-scale dependence not captured by the basis-function term. 
 The basis-function model is valid: Because $\vec{K}$ is a covariance matrix, which is a positive-semidefinite matrix, the first term in Equation \ref{eqn:C(s, u)} is positive-semidefinite, and the second term is positive-semidefinite by definition; hence, their sum is positive-semidefinite. 
 
  A classical spatial-modelling assumption is covariance stationarity and, possibly, covariance isotropy.
   \entry{Covariance stationarity}{a covariance function is stationary if it satisfies $C(\vec{s}, \vec{u}) = C^*(\vec{h})$, where the vector $\vec{h} \equiv \vec{u} - \vec{s}$ is the spatial lag between two locations $\vec{u}$ and $\vec{s}$ in $\mathbb{R}^d$.} 
   \entry{Covariance isotropy}{a covariance function is isotropic if it depends only on the distance $\norm{\vec{h}}$.}
 However, these assumptions are typically only realistic over small spatial domains. 
 In addition to guaranteeing validity, basis-function models can yield non-stationary covariance functions through the covariance matrix $\vec{K}$ or through the basis functions themselves. 
 Often, $\vec{K}$ is chosen to be of a relatively simple parametric form with minimal parameters to be estimated, in which case the burden of capturing non-stationarity falls to the basis functions.  
 
\begin{figure}
  \includegraphics[width=\linewidth]{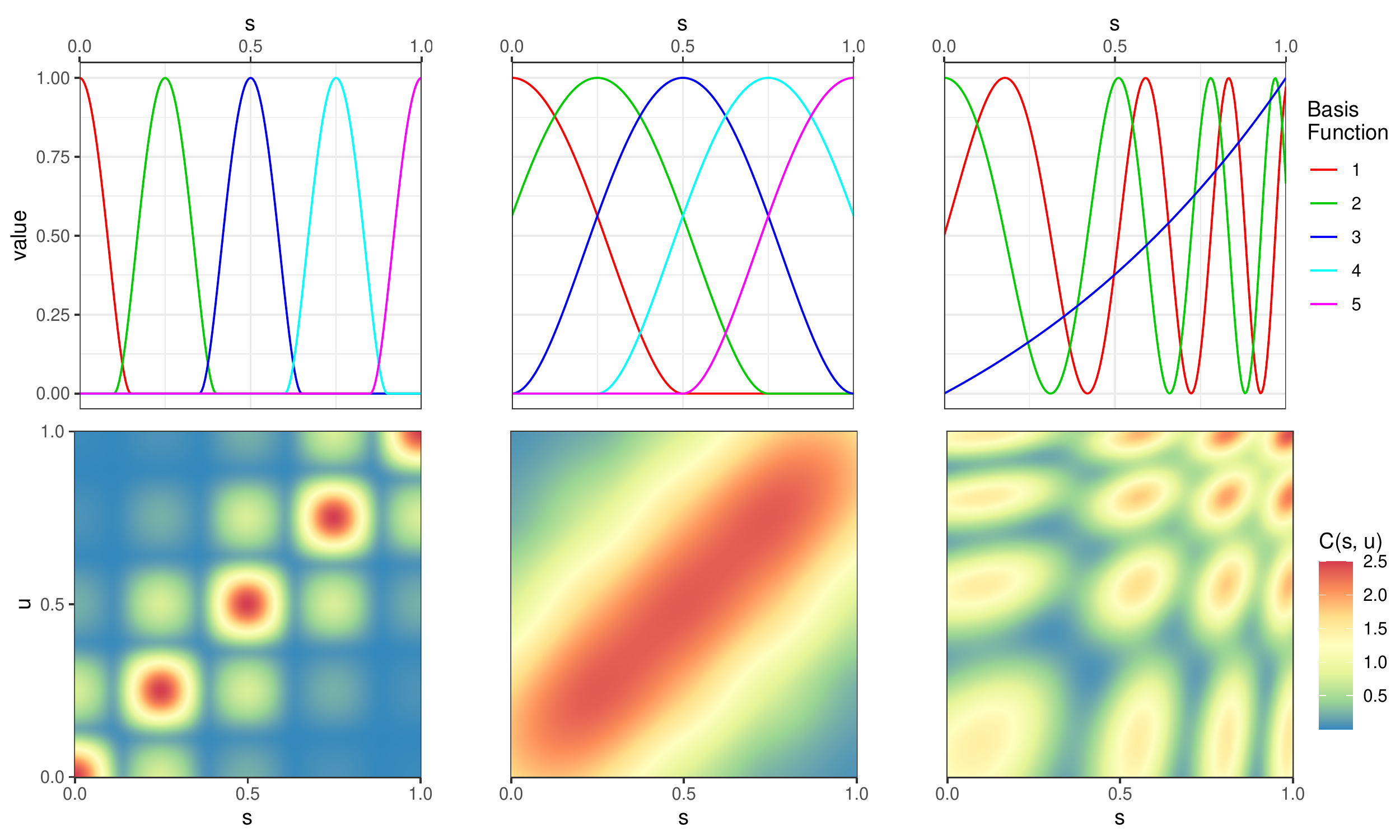}
  \caption{Three sets of bisquare basis functions (top row) on the one-dimensional spatial domain, $D = [0, 1]$, and the associated covariance functions that arise (bottom row). The first two sets of basis-function coefficients follow a first-order autoregressive model, while the third set of basis-function coefficients are modelled as independent.}
  \label{fig:basis_functions_and_implied_covariance_functions}
\end{figure}

 To illustrate this point, \textbf{Figure \ref{fig:basis_functions_and_implied_covariance_functions}} shows three sets of basis functions on the one-dimensional spatial domain, $D = [0, 1]$, and the associated covariance functions that arise when using a simple covariance matrix $\vec{K}$.
The first two sets of basis functions consist of regularly-located bisquare basis functions that have compact support; the first set has (almost) non-overlapping basis functions, while the second set has basis functions with significant overlap.
For these two sets of basis functions, $\vec{K}$ was set to be the covariance matrix of a simple first-order autoregressive process.
The non-overlapping bisquare basis functions produce undesirable artefacts in the covariance function; the overlapping bisquare basis functions yield a `pseudo-stationary' covariance function.
The third set of basis functions are highly irregular, and they are analogous to physically motivated basis functions, such as elevation, or the output of an atmospheric transport model.
Here, $\vec{K}$ was simply set to the identity matrix. Despite $\vec{K}$ having a simple structure in this third set, the resulting covariance function is clearly non-stationary. 
 This example shows two important results: First, including a degree of overlap when using basis functions with local scope can reduce undesirable artefacts in the covariance function; and second, the covariance function can be non-stationary even if $\vec{K}$ has a simple parametric form, provided the basis functions are irregular in some way. 
 In Section \ref{sec:warping_spatial_locations}, we review a recent approach that warps the spatial domain $D$ and allows highly non-stationary covariance functions to be constructed using simple structures for $\vec{K}$ \textit{and} simple basis functions.
 
For stationary covariance functions $C^{*}(\vec{h})$, continuity and differentiability properties as $\vec{h} \to \vec{0}$ are directly related to the smoothness properties of the sample paths of the underlying spatial process
(e.g., \citeauthor{Cressie_1993_stats_for_spatial_data} \citeyear{Cressie_1993_stats_for_spatial_data}, p.~60;  \citeauthor{Stein_1999} \citeyear{Stein_1999}).
These are important for determining which parameters of $C^{*}(\cdot)$ can be estimated consistently \citep{Stein_1999}. Since basis-function models are non-stationary, these theoretical results are not applicable here. Under quite restrictive assumptions applied only in one-dimensional space, \cite{Stein_2014_limitations_low_rank_approximations} critiqued spatial-basis-function models. He assumed that the true spatial model is a periodic stationary process, which is then approximated by a basis-function model that results from the truncation of the spectral representation of $\cov(\vec{Z}, \vec{Z})$. That is, he assumed that the basis functions are the orthonormal eigenvectors of $\cov(\vec{Z}, \vec{Z})$, and he added a white-noise error process to preserve the variability of the true stationary model. 
While interesting in its own right, this set-up differs substantially from that used in this review, where the true model is a non-stationary basis-function model that uses any basis functions (orthonormal or not) defined over all $\vec{s}\in D \subset \mathbb{R}^d$ for $d \geq 1$. 

There is a need for more general asymptotic results that would identify which parameters in $\vec{K}$ can be estimated consistently. Here, conditions on the smoothness (or roughness) of the basis functions will be critical, since they determine the smoothness of the underlying non-stationary process $Y(\cdot)$ in Equation 5 and, as discussed above, smoothness is key (when $Y(\cdot)$ is stationary).

 In addition to facilitating non-stationary covariance functions, it will be shown in the following section that the quadratic form in Equation \ref{eqn:C(s, u)} results in substantial computational savings when conducting spatial-statistical inference, particularly optimal spatial prediction.
 These computational savings make basis-function models well placed to handle  the `big data' often encountered in modern spatial statistics.

\section{Gaussian and non-Gaussian basis-function models}\label{part2}

Section \ref{part2} is divided into Sections \ref{sec:Basis-function representations in Gaussian spatial processes}, \ref{sec:Basis-function representations in non-Gaussian spatial processes}, and \ref{sec:Inference in a hierarchical model}, which describe well established methodologies using basis-function representations. 
 In Sections \ref{sec:Basis-function representations in Gaussian spatial processes}  and \ref{sec:Basis-function representations in non-Gaussian spatial processes}, we review basis-function representations in univariate Gaussian and non-Gaussian processes, respectively, and we consider spatial prediction of the underlying process $Y(\cdot)$. 
 In Section \ref{sec:Inference in a hierarchical model}, we discuss the empirical and Bayesian approaches to inference on the parameters of the basis-function model presented in Section \ref{sec:SpatStochasticProcceses}, along with their implications for spatial prediction of $Y(\cdot)$.

\subsection{Basis-function representations in Gaussian spatial processes}\label{sec:Basis-function representations in Gaussian spatial processes}

 In this subsection, we assume that both the process model and the data model are Gaussian. 
 \entry{Gaussian process}{$\{Y(\vec{s}): \vec{s} \in D\}$ is a Gaussian process if $\sum_{i=1}^{m}Y(\vec{s}_i)a_{i}$ is a Gaussian random variable for any $\{\vec{s}_{i}: i = 1, \dots, m\} \in D$, any real $\{a_{i}: i = 1, \dots, m\}$, and any integer $m > 0$.}
 This so-called `Gau-Gau' case is the most straightforward to work with in spatial statistics, since the likelihood, predictive means, and predictive variances are all derived from Gaussian distributions whose mean vector and covariance matrix are known. 
 However, with increasing sample sizes in many applications, direct computation of the means and covariances has become increasingly problematic. 
 The difficulty arises from the inversion of the $n\times n$ matrix $\vec{C}_Z \equiv \cov(\vec{Z}, \vec{Z})$, the covariance matrix of the spatial data $\vec{Z} \equiv (\vec{Z}(\vec{s}_1), \dots, \vec{Z}(\vec{s}_n))^\tp$ which, in general, is $O(n^3)$ in computational complexity. 
 However, when the process model has a representation in terms of $r$ basis functions, it will be seen below that inverting $\vec{C}_Z$ is generally $O(nr^2)$ in computational complexity; when $r$ is fixed, this results in $O(n)$ complexity.

\subsubsection{Gaussian data and Gaussian process (Gau-Gau) model}
 
 It will be convenient to write the model in terms of vectors and matrices. 
 Define $\vec{Y}$, $\vec{\delta}$, and $\vec{\epsilon}$ as we defined $\vec{Z}$ in the introduction to this subsection, that is, as vectors of $Y(\cdot)$, $\delta(\cdot)$, and $\epsilon(\cdot)$ respectively, evaluated at each observation location $\{\vec{s}_i: i = 1, \dots, n\}$.  
Define the $n\times p$ matrix 
 $\vec{X} \equiv (\vec{x}(\vec{s}_1), \dots, \vec{x}(\vec{s}_n))^\tp$ and the $n\times r$ matrix 
 $\vec{\Phi} \equiv (\vec{\phi}(\vec{s}_1), \dots, \vec{\phi}(\vec{s}_n))^\tp$ whose $i$th rows respectively consist of the regression covariates and the spatial basis functions evaluated at $\{\vec{s}_i: i = 1, \dots, n\}$. 
 Then the Gau-Gau model is 
 \begin{gather}
	\vec{Z} = \vec{Y} + \vec{\epsilon},\label{eqn:Gaussian_data_vector}\\
	\vec{Y} = \vec{X}\vec{\beta} + \vec{\Phi}\vec{\alpha} + \vec{\delta},\label{eqn:Gaussian_process_vector}
 \end{gather}
where $\vec{\beta}$ is a $p$-dimensional vector of fixed but generally unknown regression coefficients, $\vec{\alpha}$ is an $r$-dimensional Gaussian random vector of basis-function coefficients, $\vec{\delta}$ is an $n$-dimensional Gaussian random vector of process-model errors, and $\vec{\epsilon} \equiv (\epsilon_1, \dots, \epsilon_n)^\tp$ is an $n$-dimensional random vector of data-model errors (i.e., measurement errors) independent of $\vec{Y}$.  
 Parameters of this Gau-Gau model include $\vec{\beta}$, $\vec{K} \equiv \cov(\vec{\alpha}, \vec{\alpha})$, $\vec{C}_{\delta} \equiv \cov(\vec{\delta}, \vec{\delta})$, and $\vec{C}_{\epsilon} \equiv \cov(\vec{\epsilon}, \vec{\epsilon})$.  
 We discuss inference on the parameters in Section \ref{sec:Inference in a hierarchical model}.

\begin{figure}
  \includegraphics[width=0.75\linewidth]{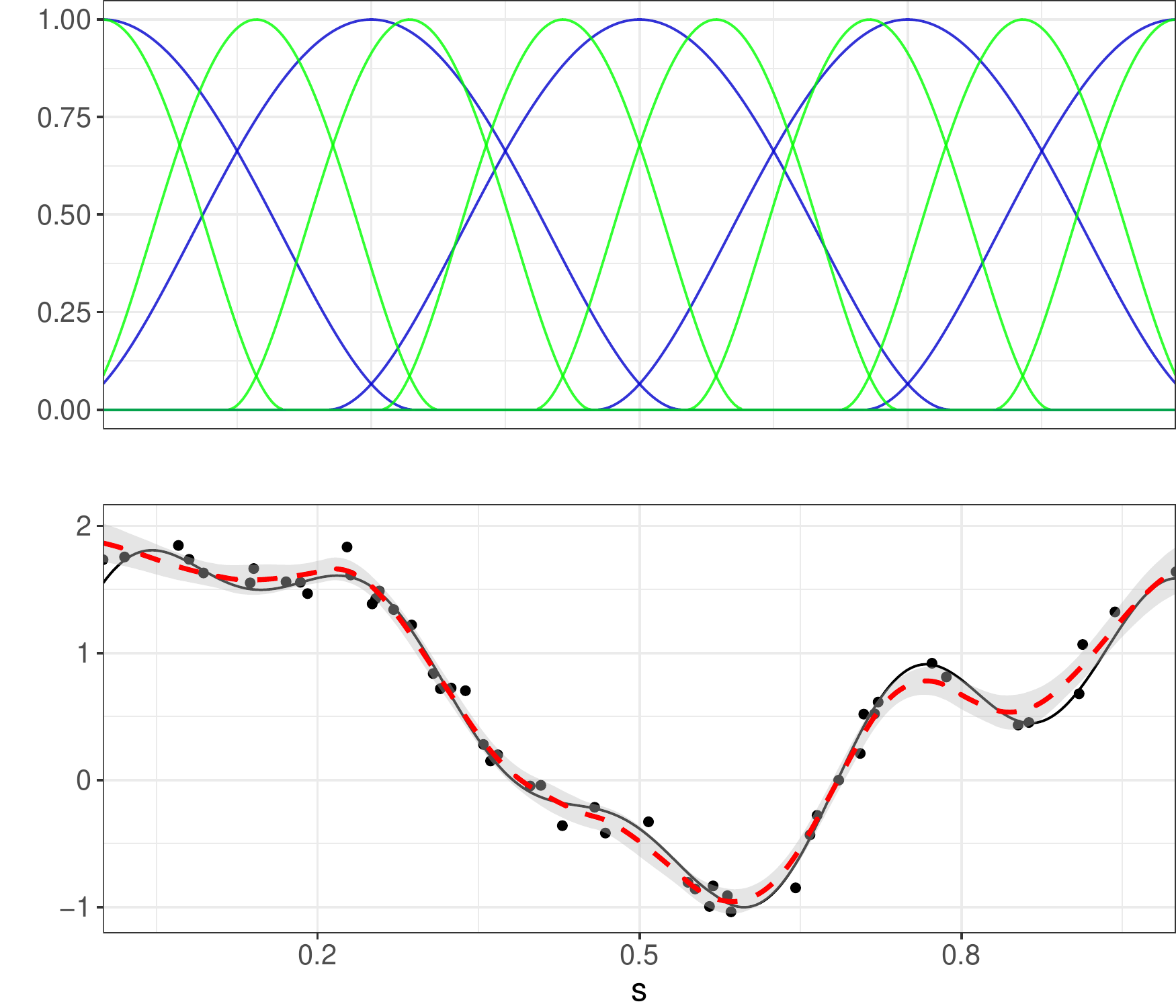}
  \caption{A simple illustration of the Gau-Gau basis-function model in Equations \ref{eqn:Gaussian_data_vector} and \ref{eqn:Gaussian_process_vector}. (Top) $r = 13$ bisquare basis functions, five of which are at a coarse resolution (blue lines) while eight are at a finer resolution (green lines). (Bottom) Hidden process $Y(\vec{s})$ for $\vec{s} \in [0, 1]$ (thin black line), data $\vec{Z}$ (black dots), optimal predictor $Y^*\!(\vec{s}_0) \equiv \E(Y(\vec{s}_0) \mid \vec{Z})$, for $\vec{s}_0 \in [0, 1]$ (red dashed line) and the 90\% pointwise prediction intervals based on the predictive standard deviation $\{\var{(Y(\vec{s}_0) \mid \vec{Z})}\}^{1/2}$, for $\vec{s}_0 \in [0, 1]$ (grey shaded region).}\label{fig:1D_toy_basis_function_example} 
\end{figure}

 The components of Equations \ref{eqn:Gaussian_data_vector} and \ref{eqn:Gaussian_process_vector} are shown for $D = [0, 1] \subset \mathbb{R}^1$ in Figure \ref{fig:1D_toy_basis_function_example}. The top of the figure shows: bisquare spatial basis functions, $\phi_1(\cdot), \dots, \phi_{13}(\cdot)$, for two resolutions at equally spaced locations within a resolution. The bottom of the figure shows: the process $Y(\vec{s})$ for $\vec{s} \in D$ (thin black line), generated by Equation \ref{eqn:general_basis-function_model} and used to construct $\vec{Y}$ in Equation \ref{eqn:Gaussian_process_vector}, with $\vec{\beta} = \vec{0}$, $\vec{\alpha} \sim \Gau(\vec{0}, \vec{K})$ for $\vec{K}$ defined by an exponential covariance function, and $\vec{\delta} = \vec{0}$; the data $\vec{Z}$ (black dots) generated by Equation \ref{eqn:Gaussian_data_vector} with $\vec{\epsilon} \sim \Gau(\vec{0}, \sigma^2_\epsilon \vec{I})$; and the optimal predictor $Y^*\!(\vec{s}_0) \equiv \E(Y(\vec{s}_0) \mid \vec{Z})$, for $\vec{s}_0 \in D$ (red dashed line) and the 90\% pointwise prediction intervals based on the predictive standard deviation $\{\var{(Y(\vec{s}_0) \mid \vec{Z})}\}^{1/2}$, for $\vec{s}_0 \in D$ (grey shaded region), obtained from Equation \ref{eqn:predictive_distribution_Gaussian} below.

 Gau-Gau spatial-basis-function models are ubiquitous in the literature; see, for example, \citet{Cressie_Johannesson_2008_FRK}, \citet{Finley_2009}, \citet{Wikle_2010_low_rank_representations_for_spatial_processes}, \citet{Lindgren_2011}, \citet{Sang_2012_approximations_of_covariance_functions}, \citet{Katzfuss_2013}, \citet{Nychka_2015}, \citet{Tzeng_2018}, and \citet{Zammit_2020}. 
 
\subsubsection{Spatial prediction}\label{sec:Gaussian_spatial_Prediction}
 
 For spatial prediction, the quantity of interest is the predictive distribution of the underlying Gaussian spatial process given the data. 
 Hence, we consider $[Y(\vec{s}_0) \mid \vec{Z}]$ for each prediction location $\vec{s}_0 \in D$. Note that prediction of $Y(\cdot)$ over subsets of $D$ is not considered here. 
 Recall from Equation \ref{eqn:predictiveDistribution} that the predictive distribution is given by ${[Y(\vec{s}_0) \mid \vec{Z}]  \propto 
 [\vec{Z} \mid Y(\vec{s}_0)][Y(\vec{s}_0)]}$, which is the product of two Gaussian distributions. By `completing the square,' it is straightforward to see that $[Y(\vec{s}_0) \mid \vec{Z}]$ is also Gaussian. 
 Computing the predictive mean and the predictive variance of $[Y(\vec{s}_0) \mid \vec{Z}]$ is perhaps most easily performed by considering the joint distribution,
\begin{equation}\label{eqn:joint-distribution_Gau-Gau}
 \begin{bmatrix}
 Y(\vec{s}_0) \\
 \vec{Z}
 \end{bmatrix}
 \sim
 \Gau\left(
  \begin{bmatrix}
 \vec{x}(\vec{s}_0)^\tp \vec{\beta} \\
 \vec{X}\vec{\beta}
 \end{bmatrix}, 
 \begin{bmatrix}
 c(\vec{s}_0) & \vec{c}_0^\tp \\
 \vec{c}_0 & \vec{C}_Z
 \end{bmatrix}
 \right), 
\end{equation}  
where $c(\vec{s}_0) \equiv \var(Y(\vec{s}_0))$, $\vec{c}_0^\tp \equiv \cov(Y(\vec{s}_0), \vec{Z}) = \cov(Y(\vec{s}_0), \vec{Y})$, and $\vec{C}_Z \equiv \cov(\vec{Z}, \vec{Z})$. 
 The predictive mean vector and the predictive covariance matrix are obtainable in closed form; as a result,
 \begin{equation}\label{eqn:predictive_distribution_Gaussian}
 Y(\vec{s}_0) \mid \vec{Z}
 \sim
 \Gau(
 \vec{x}(\vec{s}_0)^\tp \vec{\beta} + \vec{c}_0^\tp \vec{C}_Z^{-1}(\vec{Z} - \vec{X} \vec{\beta}), \;
 c(\vec{s}_0) - \vec{c}_0^\tp \vec{C}_Z^{-1} \vec{c}_0
 ). 
\end{equation} 

 The optimal predictor of the hidden value $Y(\vec{s}_0)$ is $Y^*\!(\vec{s}_0) \equiv \E(Y(\vec{s}_0) \mid \vec{Z})$ given by Equation \ref{eqn:predictive_distribution_Gaussian}. 
 That is, 
 \begin{equation}\label{eqn:Optimal_predictor}
 Y^*\!(\vec{s}_0) = \vec{x}(\vec{s}_0)^\tp\vec{\beta} + \vec{c}_0^\tp \vec{C}_Z^{-1} (\vec{Z} - \vec{X}\vec{\beta}).
 \end{equation}
 The $n\times n$ matrix, $\vec{C}_Z^{-1}$, appears in the expression for both the mean and the variance of the predictive distribution in Equation \ref{eqn:predictive_distribution_Gaussian}, and it represents a potential computational bottleneck when the sample size is large. 
 Indeed, even in the classical stationary model, which assumes a stationary covariance function for $Y(\cdot)$, $\vec{C}_Z$ generally does not have a special structure that can be exploited for computational savings, and the full cost of an $O(n^3)$ matrix inversion must be met. 
 This renders traditional spatial-prediction techniques computationally infeasible for large data sets. 
 However, there are considerable computational efficiencies afforded by spatial-basis-function models.

The basis-function model given by Equation \ref{eqn:Gaussian_process_vector} results in the following expressions for the quantities in Equations \ref{eqn:joint-distribution_Gau-Gau} and \ref{eqn:predictive_distribution_Gaussian}: 
$c(\vec{s}_0) = \vec{\phi}(\vec{s}_0)^\tp \vec{K} \vec{\phi}(\vec{s}_0) + C_\delta(\vec{s}_0, \vec{s}_0)$, $\vec{c}_0^\tp =  \vec{\phi}(\vec{s}_0)^\tp \vec{K} \vec{\Phi}^\tp + \cov(\delta(\vec{s}_0), \vec{\delta})$ and, after marginalisation,
\entry{Marginalisation}{The operation of `integrating out' random variables, in this case the random effects, in a probability model.}
\begin{equation}
\vec{C}_Z = \vec{\Phi}\vec{K} \vec{\Phi}^\tp + \vec{C}_\delta + \vec{C}_\epsilon.\label{eqn:C_Z}
\end{equation}
The optimal predictor $Y^*\!(\vec{s}_0)$ for the basis-function model is obtained by substituting these quantities into Equation \ref{eqn:Optimal_predictor}.

Typically, $\vec{C}_\epsilon$ is a diagonal matrix since it corresponds to independent measurement error, while $\vec{C}_\delta$ is often assumed to be diagonal, corresponding to fine-scale variation that is almost spatially uncorrelated \citep[see][for an exception]{Ma_2020}; here, for simplicity, we assume that $\vec{C}_\delta$ is diagonal. Now write $\vec{\xi} \equiv \vec{\delta} + \vec{\epsilon}$ and, because of the independence implied by the HM structure, $\vec{C}_\xi = \vec{C}_\delta + \vec{C}_\epsilon$, which is also a diagonal covariance matrix.
 Then Equation \ref{eqn:C_Z} may be written as
 \begin{equation}
\vec{C}_Z = \vec{\Phi}\vec{K} \vec{\Phi}^\tp + \vec{C}_\xi,\label{eqn:C_Z_II}
\end{equation}
 where $\vec{C}_\xi$ is diagonal with positive entries. Applying the Sherman--Morrison--Woodbury matrix identity \citep{Searle_Henderson_1981_inverse_of_sum_of_matrices} to Equation \ref{eqn:C_Z_II}, we have
 \begin{equation}
\vec{C}_Z^{-1} = \vec{C}_\xi^{-1} - \vec{C}_\xi^{-1}\vec{\Phi}(\vec{I} + \vec{K}\vec{\Phi}^\tp\vec{C}_\xi^{-1}\vec{\Phi})^{-1}\vec{K}\vec{\Phi}^\tp\vec{C}_\xi^{-1}.\label{eqn:C_Z_inverse}
\end{equation}
 Equation \ref{eqn:C_Z_inverse} involves inverting two matrices, an $r \times r$ matrix and the possibly-very-large but, importantly, easy-to-invert diagonal matrix $\vec{C}_\xi$. 
 \entry{Low-rank models}{models where the number of basis functions is (usually substantially) less than the number of data points.}
 We have already remarked that fixing $r$ to be a manageable number circumvents the so-called `big $n$' problem since the computational complexity of inverting $\vec{C}_Z$ drops from $O(n^3)$ to $O(nr^2)$. 
 Hence, for low-rank models where $r \ll n$, basis-function models afford scalable, $O(n)$, spatial predictions. 


\subsection{Basis-function representations in non-Gaussian spatial processes}\label{sec:Basis-function representations in non-Gaussian spatial processes}

Many spatial problems of interest deal with distinctly non-Gaussian data (e.g., counts, binary responses, extreme values), which give rise to non-Gaussian process models. 
In this subsection, we review basis-function models that account for non-Gaussianity.

\subsubsection{Trans-Gaussian spatial processes}\label{sec:TransGaussian}

The most straightforward way to construct a non-Gaussian spatial process is via a transformation $g(\cdot)$ of a spatial Gaussian process. 
We refer to these as \emph{trans-Gaussian spatial processes}. 
Consider a spatial process, $Y(\cdot) = \{Y(\vec{s}): \vec{s} \in D\}$ and a monotonic non-linear transformation $g(\cdot)$; if the process $W(\cdot) \equiv g(Y(\cdot))$ is Gaussian, then $Y(\cdot) = g^{-1}(W(\cdot))$ is a (non-Gaussian) trans-Gaussian process. 

Among the most popular transformations is the \textit{Box--Cox transformation}, $g_{\textrm{BC}}(\cdot)$, originally applied to spatial models by \citet{deOliveira_1997}. The transformation is given by
  $$
  g_{\textrm{BC}}(Y(\cdot)) \equiv \begin{cases} \frac{Y(\cdot)^\lambda - 1}{\lambda}, & \lambda \ne 0, \\ \log Y(\cdot), & \lambda = 0,  \end{cases}
  $$
  where $Y(\cdot)$ has been standardised to have no units and, for a given $\lambda$, $Y(\cdot)^\lambda$ is well defined. 
 Setting $\lambda = 0$ in the Box--Cox transformation yields a \emph{lognormal spatial process} provided $Y(\cdot) > 0$, while setting $\lambda = 1$ yields a conventional Gaussian spatial process. 
 
 Several other transformations have been considered to date in the context of spatial-process modelling. These include the square-root transform \citep{Johns_2003}; the power transform \citep{Allcroft_2003};  and the Tukey g-h transform \citep{Xu_2017}.  
 There are other innovative ways to construct non-Gaussian processes: For example, the process $(W(\cdot) + \lambda|U|)/\sqrt{V}$, where $\lambda > 0, U \sim \Gau(0, 1)$, $V \sim\textrm{Gamma}(\nu/2, \nu/2)$, and $\nu$ is a positive integer, is a skew-$t$ spatial process, which has been used to model skewed processes with heavy tails \citep[e.g.,][]{Tagle_2020}. 

The spatial models presented in the literature given just above did not employ a basis-function representation, although there is no reason why they could not have done so. 
 Indeed, a basis-function representation for $W(\cdot)$ is particularly attractive in this setting, since its dimension-reduction facilitates estimation and prediction for non-Gaussian processes, which is typically very computationally intensive. 
  \citet{Zammit_2016} used a low-rank representation of $g_{\textrm{BC}}(Y(\cdot))$ to model and predict methane sources $Y(\cdot)$, which are non-negative and skewed. There, each basis function was a two-dimensional rectangular function, equal to 1 over a small area, and equal to 0 otherwise. 
  That is, 0-1 basis functions were used to represent $Y(\cdot)$.

\subsubsection{Spatial generalised linear models}\label{sec:SpatGLMs}

\entry{Exponential family}{A set of probability distributions that include the Gaussian, Poisson, binomial, and gamma distributions.  Members of this family cover a range of supports and incorporate a wide variety of mean-variance relationships.
 }
The generalised linear model  \citep[GLM,][]{Nelder_Wedderburn_1972_GLM, McCullagh_Nelder_1989_GLM} was originally used to model non-spatial, non-Gaussian data. 
The data model assumes that, given their respective mean responses and in some cases given a dispersion parameter, the observations are conditionally independent and come from probability distributions in the \textit{exponential family}. 
The process model uses a monotonic, differentiable \textit{link function} to model the transformed mean response in terms of a linear combination of covariates. 
 The generalised linear mixed model (GLMM) extends the GLM by introducing random effects into the process model.


The \textit{spatial GLMM} \citep{Diggle_1998_spatial_GLMM} is defined as follows. 
Independently for data locations $\{\vec{s}_i: i = 1, \dots, n\}$,
\begin{gather}
    Z(\vec{s}_i) \mid Y(\vec{s}_i), \gamma \sim \text{EF}(Y(\vec{s}_i), \gamma), \label{eqn:basic_GLM_data_layer}\\
     g(Y(\vec{s}_i)) = \vec{x}(\vec{s}_i)^\tp \vec{\beta} + \eta(\vec{s}_i),\label{eqn:GLMM_systematic}
\end{gather}
where $Y(\cdot)$ is the `mean' parameter of the exponential family, $\text{EF}$ corresponds to a probability distribution in the exponential family with dispersion parameter $\gamma$, $g(\cdot)$ is a link function, the regression coefficients $\vec{\beta}$ are fixed but unknown parameters associated with covariates $\vec{x}(\cdot)$, and  $\eta(\cdot)$ is a mean-zero spatially dependent \textit{random} process, typically a Gaussian process. 
 \entry{MCMC}{Markov chain Monte Carlo is a class of algorithms often used to sample from an intractable (posterior) probability distribution.}
 \cite{Diggle_1998_spatial_GLMM} assumed a stationary Gaussian process for $\eta(\cdot)$ and used Markov chain Monte Carlo (MCMC) methods for inference and prediction.

 Since this seminal work, the spatial GLMM framework has been used extensively, and some authors have chosen to model the spatial process $\eta(\cdot)$ in Equation \ref{eqn:GLMM_systematic} with the basis-function statistical model
 \begin{equation*}
 \eta(\cdot) = \sum_{j = 1}^r \alpha_j \phi_j(\cdot) + \delta(\cdot).
 \end{equation*}
 In a remote-sensing, `big data' application, \cite{Sengupta_Cressie_2013_spatial_GLMM_FRK} showed the computational advantages of using the spatial GLMM framework; Equation \ref{eqn:basic_GLM_data_layer} was modelled with a gamma distribution, $g(\cdot) = \log(\cdot)$, and $g(Y(\cdot))$ in Equation \ref{eqn:GLMM_systematic} was given by the basis-function model, Equation \ref{eqn:general_basis-function_model}. 
 The basis-function coefficients were modelled as a Gaussian random vector with mean zero and an unconstrained covariance matrix. Then Laplace approximations in an expectation-maximisation algorithm (see Section \ref{sec:Inference in a hierarchical model}) resulted in maximum likelihood estimates of the unknown parameters. Finally, the estimates replaced the unknown parameters, and the empirical predictive distribution was generated using an MCMC algorithm.
 \entry{Empirical predictive distribution}{The predictive distribution of process values given the data, with parameter estimates replacing the unknown parameters.}

 \cite{Lee_2020_partitioned_domain_basis_function_non_Gaussian} took a scalable basis-function approach for modelling non-stationary, non-Gaussian spatial data using the spatial GLMM framework. 
 First, a clustering algorithm partitioned the spatial domain $D$ into disjoint subregions and then, for each subregion, a thin-plate-spline basis-function model was used independently of the other subregions. 
  Finally, the global process was represented as a weighted sum of the local processes.  
However, they did not include fine-scale variation in their local basis-function models (i.e., they put $\delta(\cdot) \equiv 0$), and they modelled the basis-function coefficients as independent. 



\subsubsection{Spatial prediction}

The quantity of interest is the predictive distribution of the hidden spatial process given the data, namely $[Y(\vec{s}_0) \mid \vec{Z}]$, where $\vec{s}_0 \in D$ is a given prediction location. 
 Bayes’ Rule can be used to obtain this distribution (see Equation \ref{eqn:predictiveDistribution}), which in this non-Gaussian setting is typically not available analytically. 
 Samples from the predictive distribution may be obtained via computational algorithms such as MCMC or via analytic approximations such as the Laplace approximation \citep[e.g.,][]{Sengupta_2016_GLMM_FRK_MODIS_data}. 
  \entry{Laplace approximation}{A method where a second-order Taylor-series approximation replaces the unobserved components of the complete-data likelihood function, with a multivariate Gaussian distribution, facilitating marginalisation.}
 Similar computational efficiencies in sampling from the predictive distribution in the Gaussian case, presented in Section \ref{sec:Gaussian_spatial_Prediction}, are afforded by basis-function representations of non-Gaussian spatial processes. 
 Software developments over the past decade, particularly in the open-source language \texttt{R} \citep{Rcoreteam_2021}, have led to many options for spatial prediction of non-Gaussian processes using basis-function representations; see Section \ref{sec:software} for an overview of currently available software.

\subsection{Inference in a hierarchical spatial statistical model (HM)}\label{sec:Inference in a hierarchical model}

In the subsections that follow, we describe two types of hierarchical model: One type (EHM) considers the parameters fixed, unknown, and to be estimated, and the other type (BHM) puts a Bayesian prior distribution on the parameters and computes a posterior distribution of the unknown spatial process and the parameters. In both cases, inference on the unknown spatial process is obtained from the predictive distribution.

\subsubsection{The EHM and the BHM}

Thus far, the presence of parameters $\vec{\theta}$ in the HM have been de-emphasised, but they are there! 
 If the parameters are treated as fixed but unknown, Equation \ref{eqn:JointDistribution} is re-written as:
 \begin{equation}\label{eqn:JointDistributionParametersFixed}
 [Y(\cdot), \vec{Z} \mid \vec{\theta}] = [\vec{Z} \mid Y(\cdot), \vec{\theta}] [Y(\cdot) \mid \vec{\theta}].
 \end{equation}
 Further, define the likelihood as 
 \begin{equation}
 L(\vec{\theta}; \vec{Z}) \equiv [\vec{Z} \mid \vec{\theta}] = \int [\vec{Z} \mid Y(\cdot), \vec{\theta}][Y(\cdot) \mid \vec{\theta}]\d Y(\cdot).\label{eqn:likelihood}
 \end{equation}
 Let $\vec{\Theta}$ denote the parameter space; then the \textit{maximum likelihood estimator (MLE)} of $\vec{\theta}$ is defined as $\hat{\vec{\theta}} \equiv \argsup_{\vec{\theta} \in \vec{\Theta}} L(\vec{\theta}; \vec{Z}).$  Inference on $\vec{\theta}$ proceeds by approximating $\var(\hat{\vec{\theta}})$ with the inverse of Fisher's information matrix. 
  \entry{Fisher's information matrix}{The \textit{score} vector is the derivative of the log-likelihood function with respect to the parameters. 
 Fisher's information matrix
 is the variance of the score vector.}
 This approach to handling the unknown parameters by obtaining an estimate $\hat{\vec{\theta}}$, yields what \cite{Cressie_Wikle_2011_stats_for_ST_data} called an \textit{empirical hierarchical model (EHM)} given by:
  \begin{align*}
 &\text{Data model:}  \qquad [\vec{Z} \mid Y(\cdot), \hat{\vec{\theta}}],\\
  &\text{Process model:} \qquad [Y(\cdot) \mid \hat{\vec{\theta}}].
 \end{align*}
  If, on the other hand, a probability distribution $[\vec{\theta}]$ is use to encode uncertainty in $\vec{\theta}$, Equation \ref{eqn:JointDistribution} is rewritten as:
  \begin{equation}\label{eqn:JointDistributionBayesian}
 [Y(\cdot), \vec{Z}, \vec{\theta}] = [\vec{Z} \mid Y(\cdot), \vec{\theta}] [Y(\cdot) \mid \vec{\theta}][\vec{\theta}].
 \end{equation}
The distribution $[\vec{\theta}]$ is sometimes referred to as the \textit{parameter model} \citep{Berliner_1996_hierarchical_Bayesian_time_series} but, most commonly, it is simply referred to as the \textit{parameter prior}. In this case, inferences on unknowns $Y(\cdot)$ and $\vec{\theta}$ are carried out using the posterior distribution obtained from Bayes' Rule:
 \begin{equation}\label{eqn:Bayes_rule}
 [Y(\cdot), \vec{\theta} \mid \vec{Z}] =  [Y(\cdot), \vec{\theta}, \vec{Z}] / [\vec{Z}],
 \end{equation}
 where the joint distribution on the right-hand side of Equation \ref{eqn:Bayes_rule} is given by Equation \ref{eqn:JointDistributionBayesian}, and $[\vec{Z}]$ is the `normalising constant'. This approach to handling the unknown parameters $\vec{\theta}$ yields the \textit{Bayesian hierarchical model (BHM)} given by
  \begin{align*}
 &\text{Data model:}  \qquad [\vec{Z} \mid Y(\cdot), \vec{\theta}],\\
  &\text{Process model:} \qquad [Y(\cdot) \mid \vec{\theta}],\\
    &\text{Parameter model:} \qquad [\vec{\theta}].
 \end{align*}

\subsubsection{Inference in EHMs and BHMs}
  
 To help with the exposition in this subsection, we replace generic $\vec{\theta}$ with $\vec{\theta} \equiv \{\vec{\theta}_D, \vec{\theta}_P\}$, where the data model and the process model are written as $[\vec{Z} \mid Y(\cdot), \vec{\theta}_D]$ and $[Y(\cdot) \mid \vec{\theta}_P]$ to emphasise that different parameters control different levels of the HM. 
 
In the EHM, inference on the unknown $Y(\cdot)$ is treated separately and differently from inference on the unknowns $\vec{\theta}_D$ and $\vec{\theta}_P$. 
 Recall that the likelihood is $L(\vec{\theta}_D, \vec{\theta}_P; \vec{Z}) \equiv [\vec{Z} \mid \vec{\theta}_D, \vec{\theta}_P]$ which, assuming $\vec{\theta}$ is fixed but unknown, is also the difficult-to-evaluate normalising constant, but now viewed as a function of the parameters. 
The MLE could be obtained directly if there were ways to evaluate and maximise $[\vec{Z} \mid \vec{\theta}_D, \vec{\theta}_P]$; however, this is rarely possible when the HM is not Gau-Gau as in Section \ref{sec:Basis-function representations in Gaussian spatial processes}. 
 One approach that circumvents this problem is the expectation-maximisation (E-M) algorithm \citep{Dempster_1977_EM_algorithm, McLachlan_2007_EM_algorithm}, which consists of an E-step followed by an M-step, and the steps are repeated in an iterative manner. 
 Specifically, the $l$th iteration is:
 \begin{equation}
    \begin{aligned}
 &\text{E-step: Calculate } \E(\log([\vec{Z}\mid Y(\cdot), \vec{\theta}_D] [Y(\cdot)\mid \vec{\theta}_P]) \mid \vec{Z}, \hat{\vec{\theta}}^{(l-1)}) \equiv q(\vec{\theta} \mid \hat{\vec{\theta}}^{(l-1)}), \\
  &\text{M-step: Find the }\vec{\theta} \text{ that maximises } q(\vec{\theta} \mid \hat{\vec{\theta}}^{(l-1)}); \text{ call this }\hat{\vec{\theta}}^{(l)}. 
 \end{aligned}
 \end{equation}
With a suitable starting value $\hat{\vec{\theta}}^{(0)}$ and a convergence criterion (usually depending on the closeness of successive parameter estimates and/or successive values of $q$ in the iteration), the E-M algorithm yields an MLE, $\hat{\vec{\theta}} \equiv \{\hat{\vec{\theta}}_D, \hat{\vec{\theta}}_P\}$. 
 It has been used to fit basis-function models in a spatial setting by \citet{Sengupta_2016_GLMM_FRK_MODIS_data} and in a spatio-temporal setting by \citet{Dewar_2009} and \citet{Katzfuss_Cressie_2011_spatio-temporal_smoothing_EM_algorithm}.

Now, from Bayes' Rule (see Equation \ref{eqn:Bayes_rule}), 
 \begin{equation}\label{eqn:predictive_dist_inference_section}
 [Y(\cdot) \mid \vec{Z}, \hat{\vec{\theta}}] \propto [\vec{Z} \mid Y(\cdot), \hat{\vec{\theta}}_D][Y(\cdot) \mid \hat{\vec{\theta}}_P].
 \end{equation}
Using an optimality criterion of minimising the mean-squared prediction error, the best predictor of $Y(\vec{s}_0)$ is $\E(Y(\vec{s}_0) \mid \vec{Z}, \hat{\vec{\theta}})$ with uncertainty quantified by, for example, $\var(Y(\vec{s}_0) \mid \vec{Z}, \hat{\vec{\theta}})$. This is implemented for spatial-basis-function models in, for example, \cite{Cressie_Johannesson_2008_FRK}, \cite{Cressie_Kang_2010_Empirical_SRE_FRK_spatial}, \cite{Sengupta_Cressie_2013_spatial_GLMM_FRK}, and \cite{Zammit_2021a}. 
 In the EHM, the variability due to $\hat{\vec{\theta}}$ is ignored; then closed-form formulas, approximations, or Monte Carlo sampling (e.g., MCMC) can be applied directly to find the predictive distribution with $\hat{\vec{\theta}}$ substituted for $\vec{\theta}$, as in Equation \ref{eqn:predictive_dist_inference_section}.

 \cite{Stein_2014_limitations_low_rank_approximations} gave results on estimation for rather specialised basis-function models. He found that maximum-likelihood estimators sometimes performed poorly, however his set-up was very different than that assumed in this review; see Section \ref{sec:covariance_functions} for further details. He also discussed spatial prediction using basis-function models, although he did not use a HM and hence predicted $Z(\vec{s}_0)$ instead of $Y(\vec{s}_0)$. His conclusions about spatial prediction were more circumspect; he suggested a simulation study be done to compare kriging (i.e., spatial) predictors from basis-function models,  with other spatial statistical predictors. \cite{Bradley_2016_comparison_of_spatial_predictors} used a validation approach to make such a comparison from small, medium, and large data sets, and the basis-function models performed very well, as they did under the common-task-framework design in \citet{Heaton_2019_comparative_study}.  
 
 
All inference for the BHM is in terms of the posterior distribution given by Equation \ref{eqn:Bayes_rule}. 
 However, the model, $Y(\cdot) = \vec{x}(\cdot)^\tp\vec{\beta} + \vec{\phi}(\cdot)^\tp\vec{\alpha} + \delta(\cdot)$, given by Equation \ref{eqn:general_basis-function_model} may have an identifiability problem when a prior distribution is put on $\vec{\beta}$, since $\vec{\alpha}$ is by default random. 
  The noisy and incomplete data $\vec{Z}$ could have trouble distinguishing fixed effects from random effects. In the context of lattice data observed on a lattice process, \cite{Griffith_2015_restricted_spatial_regression} proposed using a class of basis functions orthogonal to the column space of $\vec{X}$ (given in Equation \ref{eqn:Gaussian_process_vector}). 
  These are known as \textit{Moran's I basis functions}, in reference to the functional form of Moran's I statistic \citep{Cliff_Ord_1981_spatial_processes}. 
  This approach has been taken up by \cite{Hodges_2010}, \cite{Paciorek_2010_spatial_confounding}, \cite{Hughes_Haran_2013_reduced-rank_areal_spatial_GLMM}, and \cite{Bradley_2015}, and the use of these basis functions in Equations \ref{eqn:Gaussian_data_vector} and \ref{eqn:Gaussian_process_vector} has become known as restricted spatial regression (RSR). 
  In the geostatistical context, which is central to this review, \cite{Hanks_2015_restricted_spatial_regression} show that RSR does not necessarily alleviate the confounding of the fixed and random effects in the BHM. Since $\vec{\beta}$ is fixed in the EHM, it is less susceptible to this problem. 

As in the EHM, inference in the BHM is hindered by not knowing the normalising constant, here $[\vec{Z}]$. 
One popular approach that does not require its knowledge is MCMC; see, for example, \citet{Nychka_2002_multiresolution_spatial_wavelet}, \citet{Paciorek_2007_Fourier_basis_functions_spectralGP}, \citet{Wikle_2010_low_rank_representations_for_spatial_processes}, \citet{Kang_Cressie_2011_Bayesian_SRE_model}, \citet{Eidsvik_2012_predictive-process_non-Gaussian_integrated_nested_Laplace_approximations}, and \citet{Zammit_2020} for the use of MCMC in a spatial-basis-function-model setting. Approximate Bayesian techniques are also used in basis-function settings, two of the most popular being integrated nested Laplace approximations \citep[INLA; e.g.,][]{Rue_2009_INLA} and variational Bayes \citep[VB; e.g.,][]{Cseke_2016}. 
 In contrast to the EHM, variability due to uncertainty in $\vec{\theta}$ is explicitly accounted for in a BHM.

 \section{Multivariate, warped, and spatio-temporal basis-function models}\label{part3}

In Section \ref{part2}, we focused on univariate Gaussian and non-Gaussian spatial processes with basis-function components. 
 In this section, we present a generalisation of basis-function models to multivariate spatial processes, non-stationary univariate processes where the spatial domain is warped, and spatio-temporal processes.

\subsection{Basis-function representations in multivariate spatial models}\label{sec:multivariate}

Multivariate spatial models give more realistic representations of a complex world, where multiple processes interact and do not behave independently from one another. Basis-function models in a multivariate context have been used for modelling interacting geophysical processes in Antarctica \citep{Zammit_2015}, and for modelling dependence along the vertical direction in CO$_2$ fields \citep{Nguyen_2017}.

For illustration, we consider the mean-zero bivariate Gaussian process, $\vec{Y}(\cdot) \equiv \{\vec{Y}(\vec{s}): \vec{s} \in D\}$, where $\vec{Y}(\vec{s}) \equiv (Y_1(\vec{s}), Y_2(\vec{s}))^\tp$ for $\vec{s} \in D$; then $Y_1(\cdot) \equiv \{Y_1(\vec{s}): \vec{s} \in D\}$ is a mean-zero univariate spatial process, and likewise for $Y_2(\cdot)$. 
 \entry{Multivariate spatial Gaussian process in $D$}{A multivariate spatial process where all linear combinations of any of the processes at any of the spatial locations in $D$, are Gaussian.}
The covariance function for $Y_1(\cdot)$, $C_{11}(\cdot, \cdot)$, captures the within-process spatial dependence of $Y_1(\cdot)$, and likewise for $C_{22}(\cdot, \cdot)$.  The concurrent between-process spatial dependence is given by $C_{12}(\vec{s}, \vec{s})$ for $\vec{s} \in D$, and the ``spatially lagged'' between-process spatial dependence is given by 
\begin{equation}\label{eqn:cross-covariance-function}
 C_{12}(\vec{s}, \vec{u}) \equiv \cov(Y_1(\vec{s}), Y_2(\vec{u})), \quad \vec{s}, \vec{u} \in D.
\end{equation}  
 From Equation \ref{eqn:cross-covariance-function}, $C_{12}(\vec{s}, \vec{u}) = C_{21}(\vec{u}, \vec{s})$ for $\vec{s}, \vec{u} \in D$, although the special symmetry relation, $C_{12}(\vec{s}, \vec{u}) = C_{21}(\vec{s}, \vec{u})$ for $\vec{s}, \vec{u} \in D$, is not true in general.
 
 One cannot use any set of functions $C_{11}(\cdot, \cdot), C_{22}(\cdot, \cdot), C_{12}(\cdot, \cdot)$, and $C_{21}(\cdot, \cdot)$ to model the within and between spatial-statistical dependence. 
 Clearly, both $C_{11}(\cdot, \cdot)$ and $C_{22}(\cdot, \cdot)$ have to be positive-semidefinite, as explained in Section \ref{sec:Basis-function representations of spatial processes and covariance functions}. 
 However, the role of $C_{12}(\cdot, \cdot)$ and $C_{21}(\cdot, \cdot)$ is delicate, since they must bind the processes together in such a way that the matrix of covariance functions, 
\begin{equation}\label{eqn:multivariate_matrix_covariance_function}
\begin{bmatrix}
 C_{11}(\cdot, \cdot) & C_{12}(\cdot, \cdot)\\
 C_{21}(\cdot, \cdot) & C_{22}(\cdot, \cdot)
\end{bmatrix},
\end{equation}
is positive-semidefinite \citep{Genton_2015_cross_covariance_functions_multivariate_geostatistics, Cressie_Zammit_2016_multivariate}.

Now, modelling the multivariate spatial process $\vec{Y}(\cdot)$ using basis functions, we have
 \begin{equation}\label{eqn:Y_i}
 Y_i(\cdot) = 
 \vec{\phi}_i(\cdot)^\tp \vec{\alpha}_i + \delta_i(\cdot), \quad i = 1, 2,
 \end{equation}
 where $\{\delta_1(\cdot), \delta_2(\cdot)\}$ are independent mean-zero Gaussian processes with covariance functions  $\{C_{\delta_1}(\cdot,\cdot), C_{\delta_2}(\cdot,\cdot)\}$; $\vec{\alpha}_i \equiv (\alpha_{i1}, \dots, \alpha_{ir_i})^\tp$ and $\vec{\phi}_i(\cdot) \equiv (\phi_{i1}(\cdot), \dots, \phi_{ir_i}(\cdot))^\tp$ are $r_i$-dimensional vectors for $i = 1, 2$; and 
 \begin{equation*}
 \begin{bmatrix}
 \vec{\alpha}_1\\
  \vec{\alpha}_2
 \end{bmatrix}
 \sim 
 \Gau(
 \vec{0}, 
\vec{K}), 
\text{ for }   
 \vec{K} \equiv 
 \begin{bmatrix}
 \vec{K}_{11} &  \vec{K}_{12}\\
 \vec{K}_{21} &  \vec{K}_{22}
 \end{bmatrix}.
 \end{equation*}
 Then the elements of the $2 \times 2$ matrix in Equation \ref{eqn:multivariate_matrix_covariance_function} are given by 
 \begin{equation}\label{eqn:C_ij}
 C_{ij}(\vec{s}, \vec{u})
 = \begin{cases}
   \vec{\phi}_i(\vec{s})^\tp \vec{K}_{ij}\vec{\phi}_j(\vec{u}) + C_{\delta_i}(\vec{s},\vec{u}), & i = j,\\
      \vec{\phi}_i(\vec{s})^\tp \vec{K}_{ij}\vec{\phi}_j(\vec{u}), & i \ne j,
   \end{cases}
 \end{equation}
for $\vec{s}, \vec{u} \in D$ and $i, j = 1, 2$. Notice that the matrix of covariance functions with its elements given by Equation \ref{eqn:C_ij}, is positive-semidefinite when $\vec{K}$ is positive-semidefinite. However, the cross-covariance functions $C_{12}(\cdot, \cdot)$ and $C_{21}(\cdot, \cdot)$  may not satisfy the special symmetry relation, $C_{12}(\vec{s}, \vec{u}) = C_{21}(\vec{s}, \vec{u})$, which is a strength of multivariate spatial-basis-function models. 
 
 Basis-function models for a univariate spatial process are always valid; see Section \ref{sec:covariance_functions} where their positive-semidefiniteness is established. 
 The same is true for multivariate basis-function models defined by Equation \ref{eqn:Y_i}, as long as the matrix $\vec{K}$ is positive-semidefinite. 
 In contrast, the multivariate Matérn cross-covariance function requires special conditions on parameters in order for Equation \ref{eqn:multivariate_matrix_covariance_function} to be positive-semidefinite \citep{Gneiting_2010}.
 Another strong feature of covariance functions derived from univariate basis-function models is that they are generally non-stationary. 
 From Equation \ref{eqn:C_ij}, this important feature carries over to multivariate basis-function models. 
   
 Recall that we have previously mentioned spatial factor analytic models \citep{Christensen_Amemiya_2002_Latent_Variable_Analysis_of_Multivariate_Spatial_Data, Christensen_Amemiya_2003_multivariate_spatial_factor_analysis}. 
 These multivariate models look like basis-function models, but the factors are random spatial functions, and their coefficients are deterministic and must satisfy critical identifiability conditions; also see Section \ref{sec:ST_dynamic}.

 There is a \textit{conditional approach to multivariate spatial modelling} based on the decomposition
 \begin{equation}\label{eqn:multivariate_joint_Y1_Y2}
 [Y_1(\cdot), Y_2(\cdot)]
 =
  [Y_2(\cdot)\mid Y_1(\cdot)][Y_1(\cdot)],
 \end{equation}
 where $[Y_2(\cdot)\mid Y_1(\cdot)]$ is shorthand for $[Y_2(\cdot)\mid \{Y_1(\vec{s}):\vec{s}\in D\}]$; see \cite{Cressie_Zammit_2016_multivariate}. 
 The advantage of this conditional approach in Equation \ref{eqn:multivariate_joint_Y1_Y2}, in contrast to the joint approach in Equation \ref{eqn:multivariate_matrix_covariance_function}, is the ease of building positive-semidefiniteness conditions into the bivariate spatial model, as it only
 requires specification of two valid covariance functions, namely $C_{11}(\vec{s}, \vec{u})$ and $C_{2\mid 1}(\vec{s}, \vec{u}) \equiv \cov(Y_2(\vec{s}), Y_2(\vec{u}) \mid Y_1(\cdot))$. For basis-function models given by Equation \ref{eqn:Y_i}, the random coefficients $\vec{\alpha}_1$ and $\vec{\alpha}_2$ of the basis functions $\vec{\phi}_1(\cdot)$  and $\vec{\phi}_2(\cdot)$, respectively, have means equal to zero, and hence
 \begin{equation}
 \begin{aligned}
  \vec{\alpha}_2 \mid \vec{\alpha}_1 &\sim \Gau(\vec{A}\vec{\alpha}_1, \vec{K}_{2\mid 1}),\\
  \vec{\alpha}_1 &\sim \Gau(\vec{0}, \vec{K}_{11}),
 \end{aligned}
 \end{equation}
where  $\vec{A}$ is an $r_2 \times r_1$ matrix that describes the dependence of $\vec{\alpha}_2$ on $\vec{\alpha}_1$, and $\vec{K}_{2\mid 1} \equiv \cov(\vec{\alpha}_2, \vec{\alpha}_2 \mid \vec{\alpha}_1)$. 
 Then the iterated-covariance relationship gives
 \begin{align*}
 \var(\vec{\alpha}_2) &\equiv \vec{K}_{22} = \vec{K}_{2\mid 1} + \vec{A} \vec{K}_{11} \vec{A}^\tp,\\
 \cov(\vec{\alpha}_1, \vec{\alpha}_2) &\equiv \vec{K}_{12} = \vec{K}_{11} \vec{A}^\tp,
 \end{align*}
 and the joint $(r_1 + r_2) \times (r_1 + r_2)$ covariance matrix is 
 \begin{equation}
 \vec{K}
 \equiv
 \begin{bmatrix}
 \vec{K}_{11} &  \vec{K}_{12}\\
 \vec{K}_{21} & \vec{K}_{22}
 \end{bmatrix}
 =
 \begin{bmatrix}
 \vec{K}_{11} & \;\; \vec{K}_{11} \vec{A}^\tp\\
 \vec{A}\vec{K}_{11} & \;\;  \vec{K}_{2\mid 1} + \vec{A} \vec{K}_{11} \vec{A}^\tp
 \end{bmatrix}.
 \end{equation}

 Hence, the conditional approach to building bivariate basis-function models in spatial statistics is to model $\{\vec{K}_{ij}\}$ in Equation \ref{eqn:C_ij} indirectly through three matrix parameters, $\vec{K}_{11}$ ($r_1 \times r_1$, positive-semidefinite), $\vec{K}_{2\mid 1}$ ($r_2 \times r_2$, positive-semidefinite), and $\vec{A}$ ($r_2 \times r_1$, real matrix). 
 Contrast this with the joint approach that requires three matrix parameters $\vec{K}_{11}$, $\vec{K}_{22}$, and $\vec{K}_{12}$, but they are constrained  so that $\vec{K}$, the full $(r_1 + r_2) \times (r_1 + r_2)$ covariance matrix, is positive-semidefinite. 

 The Gaussian data model is
   \begin{equation}\label{eqn:multivariate_data_model}
  Z_i(\vec{s}_{ij})
  =
  Y_i(\vec{s}_{ij}) + \epsilon_{ij}, 
  \quad
    j = 1, \dots, n_i,~~ i = 1,2,
  \end{equation}
  where the measurement errors $\{\epsilon_{ij}\}$ are independent Gaussian random variables with mean zero. 
 Optimal prediction of $Y_2(\cdot)$, say, given bivariate data $\{\vec{Z}_1$, $\vec{Z}_2\}$, will be more precise than the univariate prediction of $Y_2(\cdot)$ given just $\vec{Z}_2$, since the bivariate basis-function model exploits the  dependence between coefficients $\vec{\alpha}_2$ and $\vec{\alpha}_1$. 
 If, in fact, $\E(\vec{\alpha}_2 \mid \vec{\alpha}_1) = \vec{0} = \E(\vec{\alpha}_2)$ (i.e., $\vec{A} = \vec{0}$), then the basis-function models for $Y_1(\cdot)$ and $Y_2(\cdot)$ are independent, and hence the extra data $\vec{Z}_1$ beyond the data $\vec{Z}_2$ does not improve prediction of $Y_2(\cdot)$.

\subsection{Warping the spatial locations}\label{sec:warping_spatial_locations}

We return to univariate basis-function models, but now the main aim is to model highly non-stationary fields. 
Using simple basis functions with local support induces sparsity in the basis-function matrix $\vec{\Phi}$, and using a simple parametric form for the covariance matrix $\vec{K}$ (or the precision matrix $\vec{K}^{-1}$) can dramatically reduce the number of parameters. 
 Exploiting this is central to what are commonly known as Fixed Rank Kriging (FRK) models \citep{Zammit_2021a}, LatticeKrig models \citep{Nychka_2015}, and stochastic partial differential equation (SPDE) models \citep{Lindgren_2011}; further discussion of these models is given in Section \ref{sec:software}. 
However, as seen in \textbf{Figure \ref{fig:basis_functions_and_implied_covariance_functions}}, this approach may not allow substantial spatial non-stationarity to be captured. 
 
A straightforward way to model highly non-stationary fields with basis-function models is first to \emph{warp} the domain and then to construct a spatial-basis-function model with $\vec{K}$ having a simple structure on the warped domain. In the seminal work of \citet{Sampson_1992}, warping was done using multi-dimensional scaling and thin-plate splines. 
The central idea behind warping is as follows: Suppose that a spatial process has a highly non-stationary covariance function $C_{D_G}(\vec{s}, \vec{u})$ on the original, \emph{geographic}, domain $D_G$. Then assume the existence of a warping function $\vec{f}(\cdot):D_G \rightarrow D_W$ that maps from the geographic domain $D_G$ to the \textit{warped} (or deformed) domain $D_W$, such that $C_{D_G}(\vec{s}, \vec{u}) = C_{D_W}(\vec{f}(\vec{s}), \vec{f}(\vec{u})),$ where typically $C_{D_W}(\cdot,\cdot)$ is a stationary, isotropic, covariance function of known parametric form on the warped domain $D_W$. 

There are several ways in which one can warp space, but we focus here on basis-function models. Note that spatial warping functions are necessarily multivariate since they have a multi-dimensional output (e.g., two spatial coordinates in $\mathbb{R}^2$). To the best of our knowledge, the earliest basis-function warping models were due to \citet{Smith_1996}, who modelled $\vec{f}(\cdot)$ using radial basis functions, and \citet{Perrin_1999}, who considered multiple radial basis functions linked through composition. More recently, a class of machine-learning models has emerged, known as manifold Gaussian processes, where the warping is carried out using a deep neural network. While neural networks can be treated as basis-function models, they tend to exhibit pathologies in applications in low dimensions, such as in spatial settings \citep[e.g.,][]{Dunlop_2018}. The \textit{deep compositional spatial model} of \citet{Zammit_2021b} circumvents this by forcing the mapping in each layer of the deep neural network to be injective, which precludes the possibility of `space-folding', where space is mapped onto itself and can create undesirable spatial-prediction artefacts \citep{Schmidt_2003}. These deep models bear strong connections to deep Gaussian processes \citep[e.g.,][]{Damianou_2013, Hensman_2014}, where the warping layers are modelled as (typically) low-rank Gaussian processes.

\begin{figure}
  \includegraphics[width=0.66\linewidth]{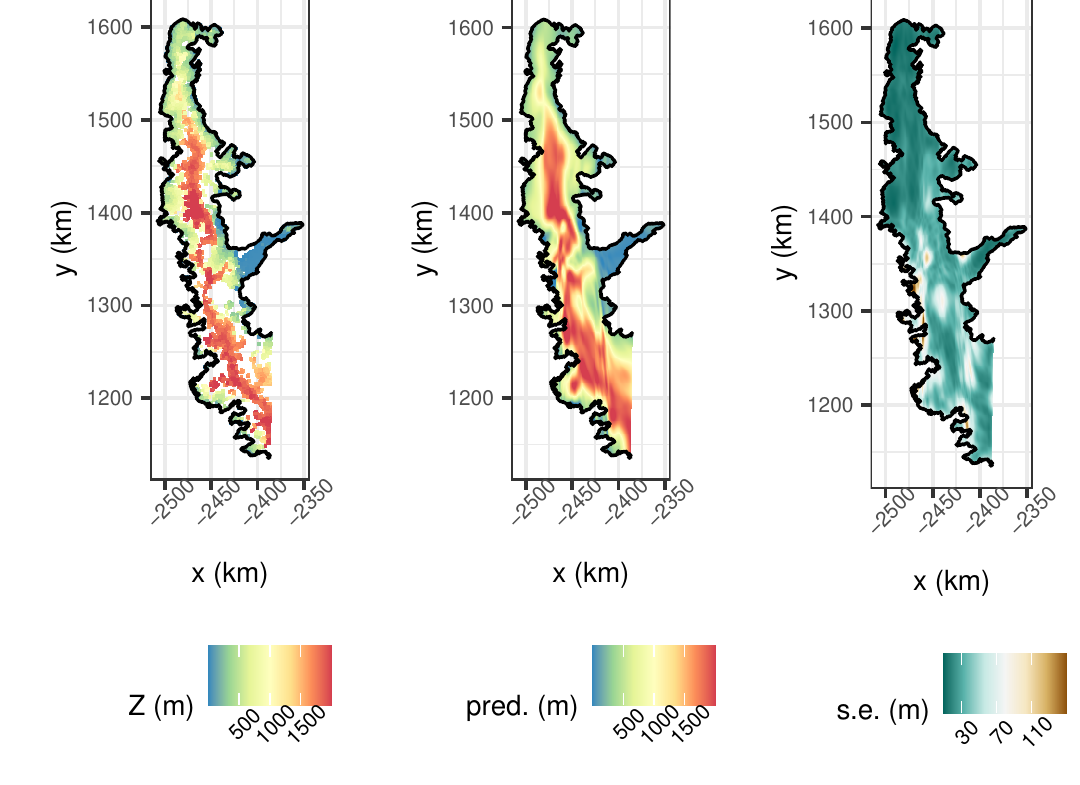}
  \caption{Spatial prediction using basis-function models of elevation (in metres) in the Northern Antarctic Peninsula. (First panel from the left) CryoSat-2 data on elevation between 2010 and 2017. (Second panel) Prediction using \texttt{deepspat}. (Third panel) Prediction standard errors obtained using \texttt{deepspat}. Only data and predictions for the Antarctic mainland (black line) are shown.\label{fig:warping}}
\end{figure}

To show the potential benefit of this type of warping using injective mappings with a basis-function component, consider elevation data from the CryoSat-2 instrument over the Northern Antarctic Peninsula. Elevation data is important when assessing the behaviour of an ice sheet, as subsequent losses or gains in observed elevation generally indicate ice-sheet-mass loss or gain, which leads to sea-level rise or fall, respectively. 
In \textbf{Figure~\ref{fig:warping}}, first panel, we show incomplete and noisy CryoSat-2 elevation data on this part of the Antarctic mainland between 2010 and 2017. 
The second panel in \textbf{Figure~\ref{fig:warping}} shows predicted elevation over the whole Northern Antarctic Peninsula using the \texttt{R} package \texttt{deepspat}\footnote{\url{https://github.com/andrewzm/deepspat}} \citep{Zammit_2021b}, which implements the deep compositional spatial model with a basis-function component. The third panel shows the prediction standard errors. 
Note how the predictions and prediction standard errors display a certain `distortion' that is not typically obtainable from spatial statistical models whose covariance functions on the geographic domain approximate a stationary covariance function. 
 Further examples are given in \citet{Zammit_2021b}.

\subsection{Basis-function models in spatio-temporal statistics}\label{sec:spatio-temporal}

 In this subsection, we briefly review basis-function models in spatio-temporal statistics. 
  Recall \textbf{Figure \ref{fig:mfresponse}}, which shows the output of the transport model MOZART in response to a pulse of $\mathrm{CO_2}$ flux in Northern Africa throughout the month of March 2015; in fact, this figure shows a spatio-temporal basis function. 
 We invoke a spatio-temporal HM with data and process indexed by $\vec{s}$ (space) and $t$ (time). Without loss of generality, we assume that the data are collected at times $t \in \{1, \dots, T\}$ and, at each time point, we have $n_t$ measurements collected across space; define
 \begin{equation*}
 \vec{Z}_t \equiv (Z(\vec{s}_{t1}), \dots, Z(\vec{s}_{tn_t}))^\tp, \quad t = 1, \dots, T.
 \end{equation*}
 The full data vector is then $\vec{Z} \equiv (\vec{Z}_1^\tp, \dots, \vec{Z}_T^\tp)^\tp$.  Again we focus on the Gau-Gau case, whereby both the data model and the underlying process are  Gaussian, and we simply note that the modelling of non-Gaussian spatio-temporal processes follows in an analogous fashion to that given in Section \ref{sec:Basis-function representations in non-Gaussian spatial processes} \citep[e.g.,][Sec.~4.5]{Wikle_2019}. 
 
 In the following subsections, we present a \textit{descriptive} approach (Section \ref{sec:ST_descriptive}) and a \textit{dynamic} approach (Section \ref{sec:ST_dynamic}) to spatio-temporal statistics, through the lens of spatio-temporal-basis-function models.

\subsubsection{Descriptive (joint) approach}\label{sec:ST_descriptive}

\entry{Stationarity}{A spatio-temporal covariance function is stationary if it can be written in terms of spatial and temporal lags, that is, $C(\vec{s}, \vec{s}^*; t, t^*)  =  C^*(\vec{h}; \tau),$  where $\vec{h} \equiv \vec{s}^* - \vec{s} $ and $\tau \equiv t^* - t$.} 
 \entry{Separability}{A stationary spatio-temporal covariance function is space-time separable if it can be written as $C^*(\vec{h}, \tau) = C^{(s)}(\vec{h}) C^{(t)}(\tau)$. It is guaranteed to be valid (i.e., positive-semidefinite) provided both the spatial covariance function, $C^{(s)}(\cdot)$, and the temporal covariance function, $C^{(t)}(\cdot)$, are valid.}
 \entry{Kronecker product}{
 The Kronecker product, $\vec{U} \otimes \vec{V}$, has $(i, j)$th block, $u_{ij}\vec{V}$. Then $(\vec{U} \otimes \vec{V})^{-1} = \vec{U}^{-1} \otimes \vec{V}^{-1}$.}
 \entry{Full symmetry}{A spatio-temporal covariance function that satisfies $C(\vec{s}, \vec{s}^*; t, t^*) = C(\vec{s}, \vec{s}^*; t^*, t)$ is said to be fully symmetric. All separable covariance functions are fully symmetric; the converse is not true.}  
 The \textit{descriptive approach} focuses on describing spatio-temporal dependence in a process $\{Y(\vec{s}, t): \vec{s}\in D, t = 1, 2, \dots \}$ through its mean function and its covariance function. 
 The mean function is $\mu(\vec{s}, t) \equiv \E(Y(\vec{s}, t))$, which is often modelled as $\vec{x}(\vec{s}, t)^\tp\vec{\beta}$ for spatio-temporal covariates $\vec{x}(\vec{s}, t)$. 
 The covariance function is $C(\vec{s}, \vec{s}^*; t, t^*)  \equiv  \cov(Y(\vec{s}, t), Y(\vec{s}^*, t^*))$.  The challenges associated with constructing valid, realistic spatial covariance functions are exacerbated in the spatio-temporal setting, since time is not just another dimension and needs to be treated differently to the spatial dimensions. 
 See \citet[][Ch.~4]{Wikle_2019} for a lengthier review than the one given below.

Classical spatio-temporal models assume \textit{second-order} (or \textit{weak}) stationarity in space and time. 
 A further classical assumption that is often made is  \textit{separability} between space and time.
 In the case of regular spatio-temporal data, where the data are observed at the same $n$ spatial locations at each time point, separability leads to significant computational savings, since the $nT$-dimensional covariance matrix, $\vec{C}_Z \equiv \cov(\vec{Z}, \vec{Z})$, can be written as a Kronecker product $\vec{C}_Z^{(s)} \otimes \vec{C}_Z^{(t)}$ of an $n$-dimensional covariance matrix $\vec{C}_Z^{(s)}$ and a $T$-dimensional covariance matrix $\vec{C}_Z^{(t)}$, facilitating computation of $\vec{C}_Z^{-1}$.  

 Basis-function models serve as an important tool for constructing valid, non-stationary, non-separable, non-fully symmetric, spatio-temporal covariance functions.
 In the spatial case, one uses known spatial basis functions, with associated random coefficients. 
 In a spatio-temporal setting, one may use spatial basis functions with temporally indexed random coefficients \citep[e.g.,][]{Wikle_1999}, or temporal basis functions with spatially indexed random coefficients \citep[e.g.,][]{Sanso_2008_Bayesian_spatio-temporal_models_discrete_convolution}, or simply spatio-temporal basis functions with random coefficients that are not indexed spatially or temporally \citep[e.g.,][]{Zammit_2021a}. 
 In this review, we present this latter case but note that similar comments can be made for the other two.
 In the descriptive approach, the extension of Equation \ref{eqn:general_basis-function_model} to a spatio-temporal setting is, for $\vec{s} \in D$ and $t = 1, 2, \dots$,  
 \begin{equation}
 Y(\vec{s}, t)
 =
 \vec{x}(\vec{s}, t)^\tp \vec{\beta} + \vec{\phi}(\vec{s}, t)^\tp \vec{\alpha} + \delta(\vec{s}, t),\label{eqn:spatio-temporal_general_basis-function_model_descriptive}
 \end{equation}
 where each term is defined analogously to its spatial counterpart; in particular,  $\vec{\phi}(\cdot, \cdot) \equiv (\phi_1(\cdot, \cdot), \dots, \phi_r(\cdot, \cdot))^\tp$ are spatio-temporal basis functions with associated random coefficients $\vec{\alpha}$, and $\delta(\cdot, \cdot)$ represents a spatio-temporal error process. 
  From Equation \ref{eqn:spatio-temporal_general_basis-function_model_descriptive}, the covariance function of $Y(\cdot, \cdot)$ is, for $\vec{s}, \vec{s}^* \in D$ and $t, t^* = 1, 2, \dots$,  
   \begin{align}
 C(\vec{s}, \vec{s}^*\!; t, t^*) 
 &= \vec{\phi}(\vec{s}, t)^\tp\vec{K}\vec{\phi}(\vec{s^*}, t^*) + \cov(\delta(\vec{s}, t), \delta(\vec{s}^*, t^*))\nonumber
 \end{align}
 which, in general, is non-stationary and non-separable. 

  Now consider Equation \ref{eqn:spatio-temporal_general_basis-function_model_descriptive} evaluated over the location-time pairs where observations occur. 
 In matrix-vector form, we have, $\vec{Y}  =  \vec{X}\vec{\beta} + \vec{\Phi}\vec{\alpha} + \vec{\delta},$ where each term is defined analogously to its spatial counterpart; see Equation \ref{eqn:Gaussian_process_vector}. 
 In a Gau-Gau model, the data model is, $\vec{Z} = \vec{Y} + \vec{\epsilon},$ where $\vec{\epsilon}$ denotes independent Gaussian error; see Equation \ref{eqn:Gaussian_data_vector}. 
 The covariance matrix of the data is, $\vec{C}_Z = \vec{\Phi}\vec{K}\vec{\Phi}^\tp + \vec{C}_\xi$, where recall that $\vec{C}_\xi \equiv \cov(\vec{\delta}, \vec{\delta}) + \cov(\vec{\epsilon}, \vec{\epsilon})$ (typically assumed to be a diagonal matrix). 
 In the low-rank case, the Sherman--Morrison--Woodbury matrix identity facilitates computation of the critically important matrix inverse $\vec{C}_Z^{-1}$ present in all inferences, without the need for separability and a Kronecker product.
 
 Choosing spatio-temporal basis functions is a task that is typically more difficult, and more consequential, than choosing spatial basis functions. 
 One straightforward approach is through the tensor product of a set of spatial basis functions and a set of temporal basis functions \citep[e.g.,][]{Zammit_2021a}; importantly, the process model does not exhibit the restrictive separability property between space and time when the tensor product is used. 
 \entry{Tensor product}{The tensor product of $(a_1, \dots, a_k)$ and $(b_1, \dots, b_l)$ is $(a_1b_1, \dots, a_1b_l, \dots,$ $a_kb_1, \dots, a_kb_l)$.}

\subsubsection{Dynamic (conditional) approach}\label{sec:ST_dynamic}

 In the \textit{dynamic approach}, the extension of Equation \ref{eqn:general_basis-function_model} to a (discrete-time) spatio-temporal setting is 
 \begin{equation}\label{eqn:spatio-temporal_general_basis-function_model_dynamic}
 Y_t(\vec{s}) = 
 \vec{x}_t(\vec{s})^\tp \vec{\beta} + \vec{\phi}(\vec{s})^\tp \vec{\alpha}_t + \delta_t(\vec{s}), \quad \vec{s} \in D, ~t = 1, 2, \dots, 
 \end{equation}
 where the different notation, $Y_t(\vec{s}) \equiv Y(\vec{s}, t)$, is used to reflect a spatial process evolving in time $t \in \{1, \dots, T\}$; $\vec{\phi}(\cdot)$ are known spatial basis functions associated with \textit{dynamically evolving coefficients} $\vec{\alpha}_t$; and $\delta_t(\cdot)$ is an error process that is independent in time. 
 Typically, a first-order Markov assumption on $\{\vec{\alpha}_t : t = 1, \dots, T\}$ is made, in which case the basis-function coefficients evolve according to
  \begin{equation}\label{eqn:dynamically_evolving_coefficients}
 \vec{\alpha}_t = 
 \vec{M} \vec{\alpha}_{t-1} + \vec{\omega}_t, \quad \text{for } t = 2, \dots, T,
 \end{equation}
 where $\vec{M}$ is referred to as the transition or propagator matrix; $\vec{\omega}_t \sim \Gau(\vec{0}, \vec{C}_\omega)$ is independent of $\vec{\alpha}_{t-1}$ and independent in time; and $\vec{C}_\omega$ is referred to as the innovation matrix. 
 Typically, $\vec{M}$ is chosen to be non-normal (i.e., $\vec{M}^\tp\vec{M} \neq \vec{M}\vec{M}^\tp$), so that the model may capture the phenomenon known as \textit{transient growth} \citep[e.g.,][p.~223]{Wikle_2019}. 
 \entry{Transient growth}{A process that is globally stable, but may have local periods of explosive growth, is said to exhibit transient growth.}
 
 Now consider the process model evaluated over the location-time pairs where observations occur. 
 In matrix-vector form, we have the process model,
  \begin{equation}
 \vec{Y}_{\!t}
 =
 \vec{X}_t\vec{\beta} + \vec{\Phi}\vec{\alpha}_t + \vec{\delta}_t, \quad t = 1, \dots, T,\label{eqn:spatio-temporal_general_basis-function_model_dynamic_matrix-vector}
 \end{equation}
 where each term is defined analogously to its spatial counterpart, and the coefficients $\{\vec{\alpha}_t\}$ evolve dynamically according to Equation \ref{eqn:dynamically_evolving_coefficients}.
 In a Gau-Gau model, the data model is,
 \begin{equation}\label{eqn:spatio-temporal_dynamic_data_model}
 \vec{Z}_t
 =
 \vec{Y}_{\!t} + \vec{\epsilon}_t, \quad t = 1, \dots, T,
 \end{equation}
 where $\vec{\epsilon}_t$ denotes Gaussian measurement error. 
 Various generalisations of Equation \ref{eqn:spatio-temporal_dynamic_data_model} are possible; for instance, one may relate $\vec{Z}_t$ and $\vec{Y}_{\!t}$ via a linear or non-linear mapping, such as discussed in \citet[][Sec.~7.1]{Cressie_Wikle_2011_stats_for_ST_data}. 
 Basis functions have often been used in dynamic spatio-temporal models, such as in \citet{Wikle_1999, Wikle_2001_spatiotemporal_tropical_ocean_surface_winds, Stroud_2001_Dynamic_spatiotemporal, Sahu_2005_Kalman_air_pollution, Sanso_2008_Bayesian_spatio-temporal_models_discrete_convolution, Dewar_2009, Cressie_2010, Freestone_2011, Cressie_Wikle_2011_stats_for_ST_data, Katzfuss_Cressie_2011_spatio-temporal_smoothing_EM_algorithm}, \citet{Zammit_2012b}, \citet{Bradley_2018_computationally_efficient_multivariate_ST_models_for_high-dimensional_count-valued_data}, and \cite{Bradley_2019_ST_models_for_big_multinomial_data_using_conditional_multivariate_logit_beta_distribution}.
 See \citet[][Ch.~5]{Wikle_2019} for a longer review than the one given above.

 In this review of spatio-temporal-basis-function models, we have focused on spatial models with known basis functions and random coefficients evolving in time. However, as noted in Section \ref{sec:Basis-function representations of spatial processes}, there is another type of spatial-basis-function model that considers the basis functions to be random and their coefficients fixed. 
 In the spatio-temporal setting, this type of model has been used for analysing Gaussian data \citep{Lopes_2008_spatial_dynamic_factor_analysis} and, using the GLMM framework, non-Gaussian data \citep{Lopes_2011_spatial_GLMM_reduced_rank_factor_analytic_model}.

\section{Software with an application}\label{part4}

 This section contains two subsections. In Section \ref{sec:software}, we discuss some of the spatial-statistical software currently available for fitting and predicting with basis-function models. In Section \ref{sec:software_application}, we give a geophysical application where six software packages are used for analysing the same sea-surface temperature data set. 
 
\subsection{Software for spatial-basis-function modelling}\label{sec:software}

 In this discussion of software that is available as of 2021 for spatial statistics with basis-function models, we focus primarily on implementations in the \texttt{R} language \citep{Rcoreteam_2021}.

 \texttt{FRK} \citep{Zammit_2021a} is a user-friendly \texttt{R} package for prediction of large spatial and spatio-temporal data sets. 
 In the spatial setting, the process model it uses is given by Equation \ref{eqn:general_basis-function_model}, with $\delta(\cdot)$ included. 
  A recent version of \texttt{FRK} \citep{Sainsbury-Dale_2021_FRKv2} caters for both Gaussian and non-Gaussian data models by employing a basis-function representation of the spatial GLMM; users can specify their own basis functions, and the default is the multi-resolution bisquare basis functions.  
  \texttt{FRK} uses spatial/spatio-temporal basic areal units (BAUs) to handle differing data supports (both point-referenced and areal) and any user-specified prediction regions.

 \texttt{INLA} \citep{Lindgren_2015} is a general-purpose \texttt{R} package for approximate inference with latent Gaussian models. From a spatial-statistics point of view, \texttt{INLA} is geared to efficiently model and fit random-fields that are solutions to SPDEs \citep{Lindgren_2011}. Specifically, the package uses finite elements (which are basis functions) to approximate the solutions to these SPDEs. \texttt{INLA} is able to cater for a number of non-Gaussian distributions, and it can accommodate different forms of geographically referenced data, as well as spatio-temporal data. \texttt{INLA} is a general-purpose package, not specifically designed with spatial prediction in mind; a recently developed package aimed at facilitating spatial modelling and fitting with \texttt{INLA} is \texttt{inlabru} \citep{Bachl_2019_inlabru}.

  \texttt{LatticeKrig} \citep{Nychka_2015} is a user-friendly \texttt{R} package for spatial prediction with large spatial data sets. 
 It uses Wendland basis functions (that have compact support) and a Markov assumption to construct a sparse precision matrix that describes the dependence between the coefficients of the basis functions. This results in efficient computations and potentially the use of a large number of basis functions. 
 \texttt{LatticeKrig} is limited to spatial, point-referenced, Gaussian data. Its model puts $\delta(\cdot) = 0$ in Equation \ref{eqn:general_basis-function_model}, which means that it omits fine-scale spatial variation. Consequently, the finest scale of the process is limited to the finest resolution of the basis functions used.
 
  \texttt{mgcv} \citep{Wood_2017_mgcv} can be used to fit and predict with spatial or spatio-temporal data using generalised additive models (GAMs), which rely on constructing smooth functions of the covariates. In a spatial setting, the covariates include spatial location. The software typically yields excellent computational times, and it can cater for a wide range of types of non-Gaussian data. The package is designed to handle point-referenced data. 
 
  The multi-resolutional approximation (MRA) of \citet{Katzfuss_2017_MRA_spatial} is a scalable approach based on a tree-structured partitioning of the spatial domain $D$ into blocks upon which predictive-process basis functions are defined. 
  \cite{Huang_2002_MRA} have a similar basis-function model but with 0-1 basis functions. Both exhibit spatial `blockiness'; see \cite{Tzeng_2005_MRA_blocking} for a way to remove this. 
 An advantage of these approaches is that they allow for easy parallelisation, which is not necessarily true for other dimension-reducing approaches. 
 Spatial prediction using MRA has been implemented efficiently in \texttt{C++} by \cite{Huang_2019_MRA_C++_implementation}.

  \begin{figure}
  \includegraphics[width=\linewidth]{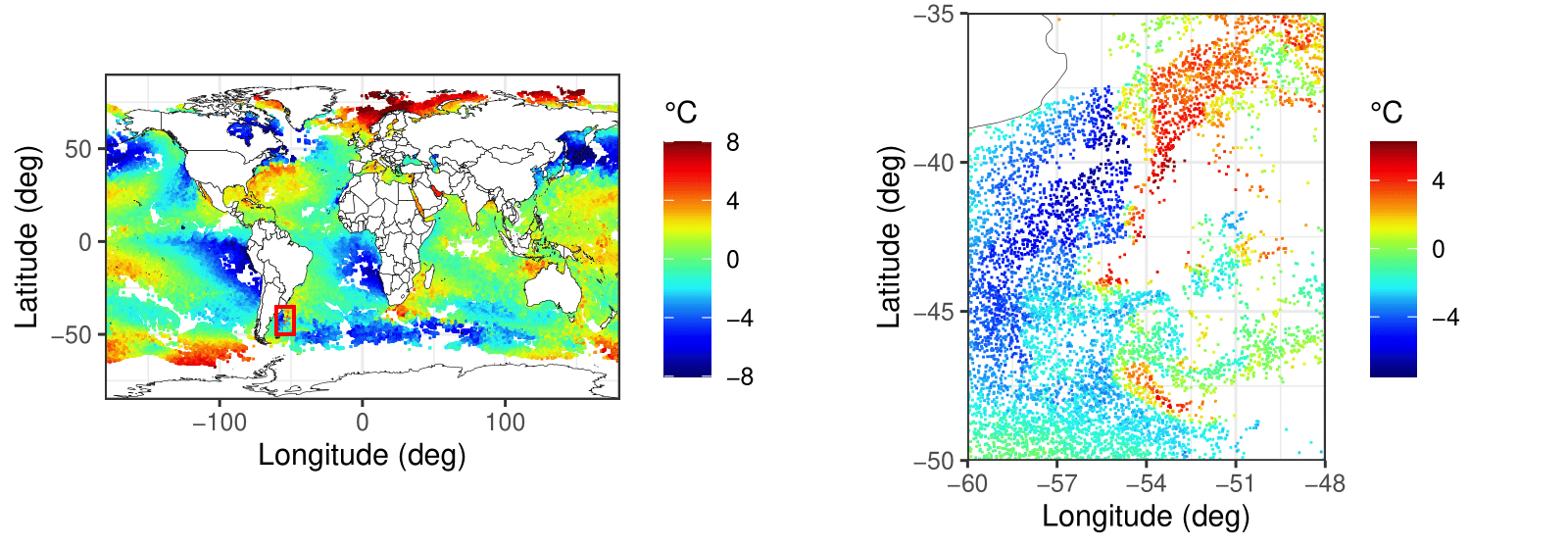}
  \caption{SST residuals used for training the models. Left panel: Global view of residuals,
with the $12^{\circ} \times 15^{\circ}$ box corresponding to the Brazil-Malvinas confluence zone marked in red. Right panel: Residuals in the Brazil-Malvinas confluence zone (note the different colour scale). Note that the testing data set is not shown, but is similar to the training set in terms of regions of data paucity and trend. }\label{fig:global_and_training_data}
\end{figure}

\subsection{Application to mapping sea-surface temperatures}\label{sec:software_application}

We now illustrate use of these software packages by way of a comparative study. 
 As a benchmark, we used traditional kriging, implemented with the \texttt{R} package \texttt{gstat} \citep{Pebesma_2004_gstat}. 
Our comparison uses the data analysed by \cite{Zammit_2020}, which consists of global sea-surface temperature (SST) data obtained from the Visible Infrared
Imaging Radiometer Suite (VIIRS) on board the Suomi National Polar-orbiting Partnership (Suomi NPP) weather satellite \citep{Cao_2013_Suomi_NPP}. 
 As in \cite{Zammit_2020}, spatial modelling was carried out on the residuals from a linear model with covariates given by an intercept, the latitude coordinate, and the square of the latitude coordinate. 
 This detrended data set is shown in \textbf{Figure \ref{fig:global_and_training_data}}, left panel. 
 In this study, we focus on a region of the ocean known as the Brazil-Malvinas confluence zone; see  \textbf{Figure \ref{fig:global_and_training_data}}, right panel.
 \entry{Brazil-Malvinas confluence zone}{An energetic region of the ocean just off the coast of Argentina and Uruguay, named after the two currents that meet at this location, namely the warm Brazil current and the cold Malvinas current.}
We divide the data within this region into a training and testing data set, each consisting of approximately 8000 observations.

The packages used in this study required several modelling decisions, which had to be made in a way that balanced predictive performance and run time. 
 We took a systematic approach to the model-selection phase by splitting the training data set in two, using half for model fitting and the other half for model validation. 
 In this way, we were able to experiment with the large number of arguments required by each package and choose the combination that was the best in terms of predictive performance and run time. 
 Typically, the most important choice for basis-function representations is the number of basis functions to use: for \texttt{FRK}, we used a total of 9210 bisquare basis functions at 4 resolutions; for \texttt{INLA}, we 
 used 11,857 basis functions; for \texttt{LatticeKrig}, we used a total of 30,482 Wendland basis functions at 4 resolutions; for \texttt{mgcv}, we used 2250 knot locations. (In \texttt{mgcv}, we used the \texttt{bam()} function, which is similar to the generalised additive model function \texttt{gam()} but optimised for large data sets.) 
 For MRA, we used 2 subregions in each partitioning, 
 49 knots in each subregion before the finest resolution, 
 and we let the software determine an appropriate number of resolutions based on the data.
 For \texttt{gstat}, we used the popular stationary Matérn covariance function, with parameters estimated by fitting the Matérn variogram to the empirical variogram.  The Matérn covariance function is $C_\nu(\vec{h}) = \sigma^2 \frac{2^{1 - \nu}}{\Gamma(\nu)} \left(\sqrt{2\nu} \frac{\|\vec{h}\|}{\rho}\right) \times$ $K_\nu \left(\sqrt{2\nu} \frac{\|\vec{h}\|}{\rho}\right)$, where $\Gamma(\cdot)$ is the gamma function, $K_\nu(\cdot)$ is the Bessel function of the second kind of order $\nu$, and $\rho$ and $\nu$ are the so-called range and smoothness parameters, respectively.

 \setlength{\textfloatsep}{0.8cm} 
 \begin{table}
\begin{center}
\caption{Comparison of spatial predictions of SST residuals in the Brazil-Malvinas confluence zone using the five \texttt{R} packages  \texttt{FRK}, \texttt{INLA}, \texttt{LatticeKrig}, \texttt{mgcv}, and \texttt{gstat}, and using a \texttt{C++} implementation of MRA  provided by \cite{Huang_2019_MRA_C++_implementation}. The diagnostics are the root mean squared prediction error (RMSPE), the empirical coverage (COV90) and the interval score (IS90) from a prediction interval with a nominal coverage of 0.9, the continuous ranked probability score (CRPS), and the total run time needed to conduct inference. Note that we omit the run time for the MRA software as it was run on a different computer than the other methods.} 
  \label{table:software_comparison}
\begin{tabular}{@{\extracolsep{5pt}} cccccc}
\\[-1.8ex] 
  \hline
  \hline
Method & RMSPE & COV90 & IS90 & CRPS & Run Time (Minutes) \\ 
  \hline
  FRK & 0.46 & 0.88 & 2.21 & 0.23 & 8.10 \\ 
  INLA & 0.46 & 0.93 & 2.18 & 0.24 & 0.86 \\ 
  LatticeKrig & 0.45 & 0.91 & 2.09 & 0.23 & 9.63 \\ 
  mgcv & 0.45 & 0.90 & 2.12 & 0.23 & 5.73 \\ 
  MRA & 0.44 & 0.92 & 2.06 & 0.23 & * \\ 
  gstat & 0.44 & 0.91 & 2.07 & 0.22 & 7.56 \\ 
   \hline
\end{tabular}
\end{center}
\end{table}

\begin{figure}
  \includegraphics[width=\linewidth]{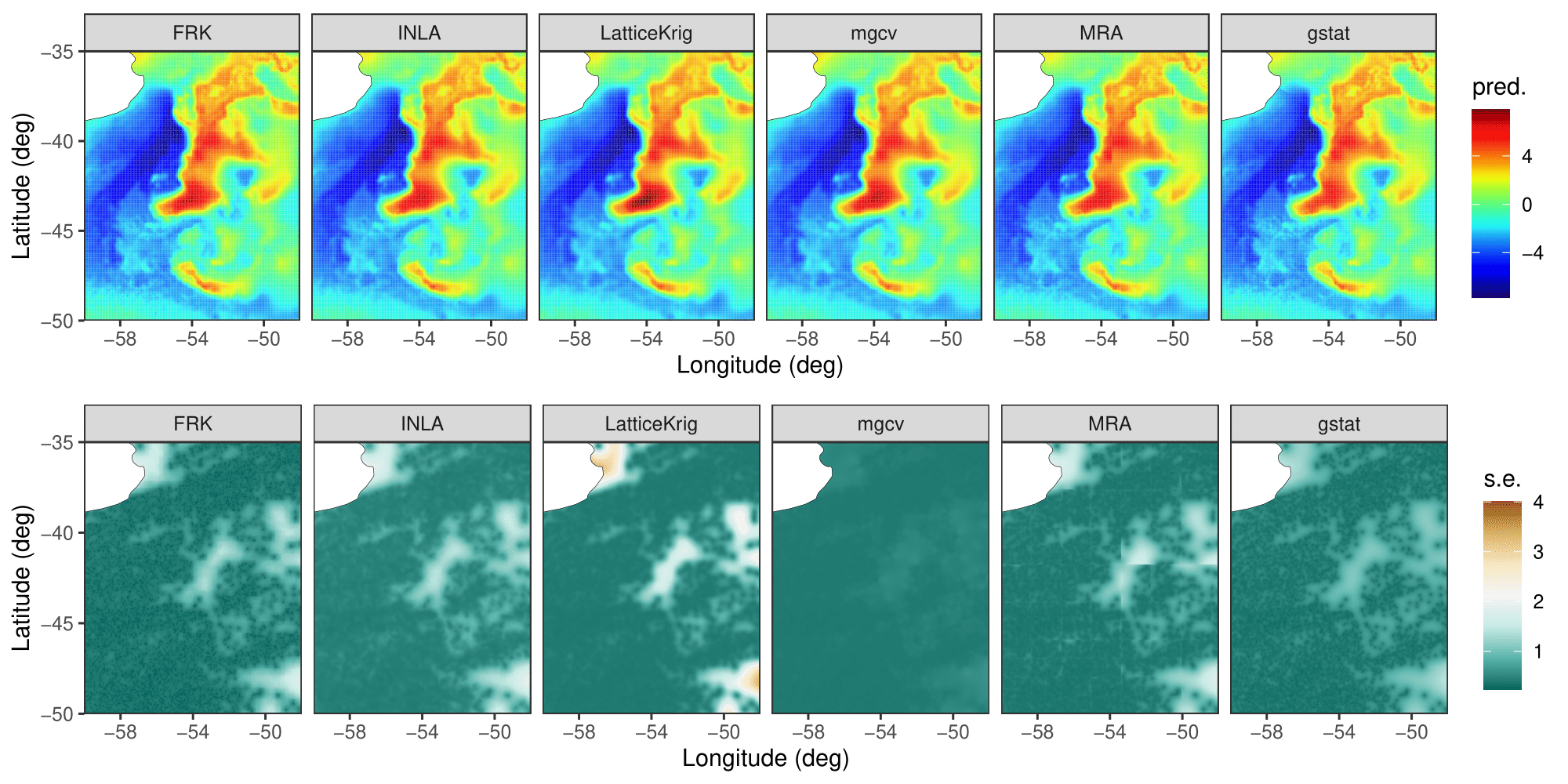}
  \caption{ 
Spatial prediction using kriging and basis-function models of SST surrounding the Brazil–Malvinas confluence zone. 
  The first row contains the spatial predictions, while the second contains the prediction standard errors.
  The first column corresponds to the results obtained using \texttt{FRK}; the second to \texttt{INLA}; the third to \texttt{LatticeKrig}; the fourth to \texttt{mgcv}; the fifth to MRA; and the sixth to \texttt{gstat}. 
  The predictions and standard errors are similar for all methods. 
   Note that blocks are visible if one ``zooms in'' on the MRA predictions or standard errors when viewing the pdf form of this article; this ``blockiness'' is a well known feature of multi-resolutional approximations. 
  }\label{fig:Brazil_Malvinas_results}
\end{figure}

 \textbf{Table \ref{table:software_comparison}} summarises the results on the testing data set.
 The results are similar across all methods; this is reinforced by  \textbf{Figure \ref{fig:Brazil_Malvinas_results}}, which shows predictions and prediction standard errors over the entire domain of interest. 
 In unobserved regions, the prediction standard errors from all packages are similar, apart from those from \texttt{mgcv}, which are smaller.
 
 Interestingly, all methods perform similarly to \texttt{gstat} on this example, which is using simple kriging for prediction of the residual process. Indeed, on data sets of sizes 10,000 or less, there is often little benefit in using basis functions for spatial prediction over kriging, unless the basis functions are specifically chosen to reproduce a non-stationary feature in the data. However, there are still benefits to the basis-function approach; in particular, if the sample size had been an order of magnitude larger, \texttt{gstat} would not have been computationally feasible (unless local kriging were used).

   Implementing from scratch the methods used in this study would have been an arduous undertaking. 
 Fortunately, thanks to the open-source software \texttt{R} and the many contributions from its community, our implementation of these complicated methods involved only a few lines of code. 
 For the code used in this study, please see \mbox{\url{https://github.com/msainsburydale/ARSIA_BasisFunctionModels_code}}.

\section{Epilogue}\label{sec:epilogue}
\addcontentsline{toc}{section}{\nameref{sec:epilogue}}

This review of basis-function models in spatial statistics has highlighted the reach and potential these models have for analysing spatial and spatio-temporal data. With technological advances in data collection, storage, and management, and with a realisation that the `Where' (and `When') question is highly relevant to answering important `Why' questions, these non-stationary and computationally scalable spatial statistical models offer a path forward. 
 We have emphasised here the importance of not only producing a parameter estimate or a spatial prediction, but also of quantifying uncertainties as part of these inferences. 
 Sections \ref{part1}--\ref{part4} establish basis-function modelling in specific areas of spatial statistics but, in principle, any area of spatial statistics that involves spatial-covariance modelling could straightforwardly incorporate basis-function models.

\section*{ACKNOWLEDGEMENTS}
Noel Cressie's and Andrew Zammit-Mangion's research was supported by an Australian Research Council (ARC) Discovery Project, DP190100180. Andrew Zammit-Mangion's research was also supported by an ARC Discovery Early Career Research Award, DE180100203. 
Matthew Sainsbury-Dale's research was supported by an Australian Government Research Training Program Scholarship. 
The authors would like to thank Ann Stavert for generating the MOZART output shown in {\bf Figure}~\ref{fig:mfresponse}; Stephen Chuter and Geoffrey Dawson for providing the processed CryoSat-2 data, originally made available by the European Space Agency, which was used in the example of Section \ref{sec:warping_spatial_locations}; Yuliya Marchetti for providing the sea-surface temperature data set in the case study of Section~\ref{sec:software_application}; and Yi Cao for running the MRA software for the example of Section~\ref{sec:software_application}. 
 We are also grateful to an anonymous reviewer whose suggestions enhanced the presentation of our article. 

\bibliographystyle{BasisFunctionModels} 
\bibliography{BasisFunctionModels}

\end{document}